\documentclass[english,nolineno]{socg-lipics-v2021}
\usepackage{graphicx}  
 
 \usepackage{thmtools} 
\usepackage{thm-restate}
\usepackage{amsmath,amsthm,stmaryrd,amssymb,amsfonts}
\usepackage{vmargin,xspace}
\newif\iflong
\newif\ifshort
\usepackage{cleveref}
\longtrue

\iflong
\else
 \shorttrue
\fi
\newtheorem{longtheorem}{Theorem}
\newtheorem{longlemma}[longtheorem]{Lemma}
\newtheorem{longproposition}[longtheorem]{Proposition}
\newtheorem{longdefinition}[longtheorem]{Definition}
\newtheorem{longobservation}[longtheorem]{Observation}
\newtheorem{redrule}{Reduction Rule}

 \newcommand{\YES}{\textup{\textsf{YES}}}

\newcommand{\Oh}{{\mathcal O}}
\newcommand{\nat}{\mathbb{N}}

\newcommand{\F}{\mathcal{F}}
\newcommand{\M}{\mathcal{M}}

\newcommand{\PSPACE}{\mbox{\sf PSPACE}}
\newcommand{\NP}{\mbox{{\sf NP}}}

\newcommand{\FPT}{\mbox{{\sf FPT}}}

\newcommand{\W}{\mbox{{\sf W}}}

\newcommand{\R}{\mathcal{R}}
\newcommand{\SSS}{\mathbf{\emph{S}}}
\newcommand{\CCC}{\mathcal{C}}
\newcommand{\dist}{\emph{\textnormal{dist}}}
  
\newcommand{\cmpl}{\textsc{\textup{CSMP}}\xspace}

 \newcommand{\blockd}{\ensuremath{\emph{blockd}}}
 \newcommand{\III}{\mathcal{I}}
 \newcommand{\planaroneagentproblem}{\textup{\textsc{Planar-CSMP-1}}\xspace}
  \newcommand{\bigoh}{{\mathcal O}}

\newcommand{\macrofk}{2^{\bigoh(k^{4})}}

\usepackage{todonotes,verbatim}

\usepackage[noadjust]{cite}

\title{A Minor-Testing Approach for
Coordinated Motion Planning with Sliding Robots}

\titlerunning{A Minor-Testing Approach for
Coordinated Motion Planning with Sliding Robots}

\author{Eduard Eiben}{Department of Computer Science, Royal Holloway, University of London, Egham, UK}{eduard.eiben@rhul.ac.uk}{https://orcid.org/0000-0003-2628-3435}{}

\author{Robert Ganian}{Algorithms and Complexity Group, TU Wien, Vienna, Austria}{rganian@gmail.com}{https://orcid.org/0000-0002-7762-8045}{Projects No.\ 10.55776/Y1329 and 10.55776/COE12 of the Austrian Science Fund (FWF), Project No.\ 10.47379/ICT22029 of the Vienna Science Foundation (WWTF)}

\author{Iyad Kanj}{School of Computing, DePaul University, Chicago, USA}{ikanj@cdm.depaul.edu}{0000-0003-1698-8829}{}

\author{Ramanujan M.~Sridharan}{Department of Computer Science, University of Warwick, UK}{R.Maadapuzhi-Sridharan@warwick.ac.uk}{ }{}

\authorrunning{Eiben et al.} 

\Copyright{Eiben et al.}
\date{}

\begin{document}

\ccsdesc[300]{Theory of computation~Parameterized complexity and exact algorithms}

\keywords{coordinated motion planning on graphs, parameterized complexity, topological minor testing, planar graphs
}

\maketitle
\begin{abstract}
We study a variant of the Coordinated Motion Planning problem on undirected graphs, referred to herein as the \textsc{Coordinated Sliding-Motion Planning} (CSMP) problem. In this variant, we are given an undirected graph $G$, $k$ robots $R_1,\dots,R_k$ positioned on distinct vertices of $G$, $p\leq k$ distinct destination vertices for robots $R_1,\dots,R_p$, and $\ell \in \nat$. The problem is to decide if there is a serial schedule of at most $\ell$ moves (i.e., of makespan $\ell$) such that at the end of the schedule each robot with a destination reaches it, where a robot's move is a free path (unoccupied by any robots) from its current position to an unoccupied vertex. The problem is known to be \NP-hard even on full grids. It has been studied in several contexts, including coin movement and reconfiguration problems, with respect to feasibility, complexity, and approximation. Geometric variants of the problem, in which congruent geometric-shape robots (e.g., unit disk/squares) slide or translate in the Euclidean plane, have also been studied extensively. 

We investigate the parameterized complexity of CSMP with respect to two parameters: the number $k$ of robots and the makespan $\ell$. As our first result, we present a fixed-parameter algorithm for CSMP parameterized by $k$. For our second result, we present a fixed-parameter algorithm parameterized by $\ell$ for the special case of CSMP in which only a single robot has a destination and the graph is planar, which we prove to be \NP-complete. A crucial new ingredient for both of our results is that the solution admits a succinct representation as a small labeled topological minor of the input graph. 
 
\end{abstract}

\section{Introduction}
% \paragraph{Motivation.}
In the coordinated motion planning problem, we are given an integer $\ell\in \nat$, an undirected graph $G$, 
$k$ robots $R_1,\dots,R_k$ positioned on distinct starting vertices of $G$ and $p\leq k$ distinct destination vertices for robots $R_1,\dots,R_p$.
%We assume that the $k$ starting vertices are distinct, and so are the $k$ destination vertices. 
In each step, one robot (or a subset of robots in other variants of the problem in which robots may move in parallel) may move from its current vertex to an adjacent unoccupied vertex. The goal is to decide if there is a schedule (i.e., a sequence of moves) of length at most $\ell$, at the end of which each robot $R_1,\dots,R_p$ has reached its destination vertex without colliding with other robots along the way. The problem and its variants (including the common variant where $p=k$) have been extensively studied, with respect to various motion types and graph classes, due to their ubiquitous applications in artificial intelligence (planning), robotics, computational geometry and more generally theoretical computer science~\cite{egksocg2023,DeligkasEGK024,heuristic3,heuristic2,heuristic1,heuristic5,heuristic4,kanjparsa,lavalle,lavalle1,halprinunlabeled,demaine,survey,alagar,sharir, sharir1, sharir2}. Moreover, it generalizes several well-known puzzles, including Sam Loyd's famous 15-puzzle---known as the ($n^2-1$)-puzzle---and the Rush-Hour Puzzle. Most of these variants are \NP-hard, including the aforementioned ($n^2-1$)-puzzle~\cite{nphard2,nphard1}. Due to its numerous applications, the coordinated motion planning problem on (full rectangular) grids was posed as the SoCG 2021 Challenge~\cite{socg2021}.

In this paper, we consider a variant of the coordinated motion planning problem on undirected graphs in which robots move serially one at a time, and a robot's move consists of a move along a free/unobstructed path (i.e., a path in which no other robot is located at the time of the move) as opposed to a single edge; we refer to such a move as a \emph{sliding} move, analogous to the type of motion in geometric variants of the coordinated motion planning problem, where robots are geometric shapes (e.g., disks or rectangles) that move by translating/sliding in the plane. These geometric variants have been extensively studied since the 1980s~\cite{alagar,kanjparsa,sharir, sharir1, sharir2}. The goal of this variant is  to decide whether there is a schedule of makespan at most $\ell$, where here the makespan of the schedule is simply the number of moves in it. 
%We will generalize the above variant to the scenario where $p \leq k$ of the robots, $R_1, \ldots, R_p$, have destinations (the case $p=k$ corresponds to the classical case where all robots have destinations). We refer to this variant in this paper as the \textsc{Coordinated Sliding-Motion Planning} (CSMP) problem.  
We denote our general problem of interest as \textsc{CSMP}. We will also consider the special case of \textsc{CSMP} where $G$ is planar and $p=1$---that is, only a designated robot has a destination and the remaining robots merely act as ``blockers''. We refer to this special case as \planaroneagentproblem.

\subparagraph{Contributions and Techniques.} We present a fixed-parameter algorithm for CSMP parameterized by the number $k$ of robots, and a fixed-parameter algorithm for \planaroneagentproblem{} parameterized by the makespan $\ell$,  after proving that \planaroneagentproblem{} remains \NP-hard. A crucial common ingredient for both results is showing that the solution to a problem instance admits a succinct representation, whose size is a function of the parameter, as a topological minor of the input graph. To the best of our knowledge, this is a novel approach in the context of coordinated motion planning problems.

\smallskip

\noindent \textbf{\cmpl.} Showing that the solution admits a succinct representation entails first bounding the makespan in an optimal solution by a function of  the number $k$ of robots. 
Towards this end, we preprocess the graph by shortening (to $\bigoh(k)$) sufficiently long paths comprising degree-2 vertices and focus on two cases. In one case, where the graph has bounded\footnote{In the rest of this description, we simply say {\em bounded} to refer to values bounded by a function of $k$.} treedepth, we show that the instance can be reduced iteratively without altering the solution. This reduction process ultimately results in an instance of bounded size, implying the existence of a solution with a bounded makespan. 

In the other case, we use ideas from \cite{DeligkasEGK024} to define a notion of {\em havens}, which are subgraphs of bounded size 
% that can efficiently handle collisions. Here, by ``efficiently handling'' collisions, we mean that any instance of {\cmpl} where the input graph is a haven, has a solution with $k^{\bigoh(1)}$ moves.\todo{I: Maybe we shouldn't describe this in terms of the input graph.\\ RMS: removed this and patched the pre and post with some additional words after that.} Thus, 
with the following property: if a robot has to pass through a haven, then with a bounded number of additional steps,  
% between the time step it enters the haven and the time step it exits,\todo{I: Maybe remove this part ``between...exits". RMS: done}  
we can enable it to pass through the haven regardless of the other robots in the haven (i.e., without collision), effectively ``untangling'' all the robots' traversals within the haven. Building on this, we construct a schedule in which each robot interacts with only a bounded number of havens, ultimately leading to a bounded-makespan schedule.
% \todo{I: Bounded could be confusing as it may suggest constant. Maybe add a footnote that bounded is w.r.t.~the parameter. \\ RMS: done. 
% }

After bounding the makespan in an optimal solution, we show that it essentially describes a graph $H$ of bounded size as a topological minor in the input graph. Here, we have to 
work with {\em rooted} topological minors since all robots have specified starting points and some have specified destinations. For us, the roots are just the starting and ending vertices, as we have boundedly-many roots. Towards our goal, we partition the vertices of the graph induced by the edges participating in an optimal solution into two sets: important and unimportant vertices.  
We show that the number of important vertices is bounded, and that the unimportant vertices form paths whose internal vertices have degree $2$. This structure enables us to show that the graph induced by the edges in the solution (with terminals as roots) is a realization of some rooted graph with a bounded number of vertices as a topological minor in the input graph.  
We also prove the converse, namely that from {\em any} realization of $H$ as a topological minor in $G$, one can obtain an optimal solution for the given instance. 
Towards proving this, we show that if a robot follows a path that is devoid of important vertices, then it slides on that path without stopping.
%every robot that visits a vertex in a path that is devoid of important vertices, simply slides along the path without stopping on any of its vertices. 
Finally, we leverage known results on topological minor containment to verify the existence of the correct rooted graph.  

\smallskip

 % For \planaroneagentproblem{},
 \noindent
{\bf\planaroneagentproblem.} It is easy to see here that the solution admits a succinct representation since the parameter itself is the makespan. The difficulty is that the number of robots in the instance can be very large compared to the parameter $\ell$, and we cannot find a realization of the succinct solution which avoids the ``blocker'' robots that do not contribute to the realization.
%seeking a realization of the succinct solution may not work since the realization may contain many ``blocker'' robots (i.e., robots other than the main designated robot with destination) along its paths. 
To cope with this hurdle, we exploit the planarity of the graph to reduce its diameter, thus reducing the instance to an instance on a graph with bounded treewidth. The fixed-parameter tractability of the problem will then follow from Courcelle's Theorem~\cite{Courcelle90,ArnborgLS91}.
 
The heart of our reduction method is finding an ``irrelevant edge'' in the graph that can be safely contracted to produce an equivalent instance of the problem. While the method of finding an irrelevant edge and contracting it has been employed in several results in the literature to exclude large grids, thus reducing the treewidth of the graph, in our setting the grid approach seems unworkable.
Instead, we show that a sufficiently-large component $C$ of free vertices (i.e., vertices with no blockers on them) must contain an irrelevant edge. We reduce all the cases to one where $C$ is formed by a skeleton consisting of a ``long'' path of degree-2 vertices (the degree is taken in $C$), plus a ``small'' number of additional vertices. We then show that if the instance is a \YES-instance of the problem, then it admits a solution that interacts nicely with this skeleton, in the sense that we can identify a subgraph of the representation of the solution, i.e., a topological minor that is essentially a roadmap of the solution, separated from the rest of the solution by a small cut. We can enumerate each possible representation of the part of the solution that interacts with the skeleton, and for each possible representation, test whether it exists near the skeleton; if it does, we mark the vertices in the graph that realize this representation. Since the skeleton is long, the total number of marked vertices on the skeleton is small, and an edge with both endpoints unmarked---and hence can be contracted---must exist.  
  
\subparagraph{Related Work.} The CSMP problem, with $p=k$, has been studied in~\cite{calinescu2008} on several graph classes, and with respect to various settings, based on whether or not the destinations of the robots are distinguished (referred to as labeled or unlabeled). The graph classes that were considered include trees, grids, planar graphs, and general graphs, and it was shown that the problem is \NP-hard even for (full rectangular) grids. Upper and lower bounds on the makespan of a solution, as well as computational complexity and approximation results, were derived for the various  settings considered in~\cite{calinescu2008}. 

A related problem to \textsc{CSMP} is that of moving unit disks (or objects~\cite{survey}) in the plane, which has been studied in the context of reconfiguring/moving coins in the plane~\cite{coins,bereg2008}, and was shown to be \NP-hard~\cite{coins}. We also mention the related work on coordinated pebble motion on graphs (and permutation groups), which was studied in~\cite{Korhonen21} motivated by the work on the ($n^2-1$)-puzzle. 

The variant of coordinated motion planning in which only one designated robot is required to reach its destination, that is $p=1$, and each move is along a single edge, was studied as \textsc{Graph Motion Planning with 1 Robot} in~\cite{PapadimitriouRST94}, where complexity and approximation results were derived. 
 
Despite the tremendous amount of work on coordinated motion planning problems, not much work has been done from the parameterized complexity perspective. %Perhaps the most relevant of this work to the problems under consideration is the work in~\cite{DeligkasEGK024,EibenGK23}. 
The work in~\cite{DeligkasEGK024} studied the parameterized complexity of the non-sliding version of \textsc{CSMP}, that is, where each motion step is along a single edge, parameterized by the makespan in the serial setting. Another recent work~\cite{egksocg2023} studied the parameterized complexity of the classical coordinated motion planning problem on grids, presenting parameterized algorithms for various objective measures (makespan, travel length).  
The work in~\cite{KnopEtalMAPF24} established the \W[1]-hardness of the classical coordinated motion planning problem parameterized by the makespan in the parallel motion setting, and also showed the \NP-hardness of this problem on trees, among other results. Finally, the parameterized complexity of a geometric variant of the \PSPACE-complete Rush-Hour problem, which itself was shown to be \PSPACE-complete~\cite{flake2002}, was studied~\cite{fernaucccg}; in particular, that problem was shown to be \FPT{} when parameterized by either the number of robots (i.e., cars) or the number of moves.   

\section{Terminology and Problem Definition}
\label{sec:prelim}

\iflong

\subparagraph*{Graphs.} 
For a graph $G$, ${\sf paths}(G)$ denotes the set of all simple paths in $G$.
The length of a path is the number of edges in it.
 For a path $P$ and vertices $u,v$ in it, we denote by $P[u,v]$ the subpath of $P$ starting at $u$ and ending at $v$.  The \emph{$P$-distance} between $u$ and $v$, denoted $\dist_{P}(u, v)$, is the length of $P[u, v]$. 
\subparagraph*{Parameterized Complexity.}
 A {\it parameterized problem} $Q$ is a subset of $\Omega^* \times
\nat$, where $\Omega$ is a fixed alphabet. Each instance of $Q$ is a 
pair $(I, \kappa)$, 
where $\kappa \in \nat$ is called the {\it
parameter}. A parameterized problem $Q$ is
{\it fixed-parameter tractable} (\FPT)~\cite{CyganFKLMPPS15,DowneyFellows13,FlumGrohe06}, if there is an
algorithm, called a {\em fixed-parameter algorithm} or an {\em \FPT-algorithm}, that decides whether an input $(I, \kappa)$
is a member of $Q$ in time $f(\kappa) \cdot |I|^{\bigoh(1)}$, where $f$ is a computable function and $|I|$ is the input instance size.  The class \FPT{} denotes the class of all fixed-parameter tractable parameterized problems.

  We  refer to~\cite{CyganFKLMPPS15,DowneyFellows13} for more information on parameterized complexity. We write $[n]$, where $n \in \nat$, for $\{1, \ldots, n\}$.
\fi

\emph{Treewidth} is a fundamental graph parameter which can be seen as a measure of how similar a graph is to a tree; trees have treewidth $1$, while the complete $n$-vertex graph has treewidth $n-1$. A formal definition of treewidth will not be necessary to obtain our results; however, we will make use of the fact that every planar graph with diameter $r$ has treewidth at most $3r+1$~\cite{RobertsonS84,biedl} and 
\iflong
of Courcelle's Theorem (introduced below). 
\fi

\iflong
 \subparagraph{Monadic Second Order Logic.} 
	We consider \emph{Monadic Second Order} (MSO) logic on edge- and vertex-labeled
	graphs in terms of their incidence structure, whose universe contains vertices and
	edges; the incidence between vertices and edges is represented by a
	binary relation. We assume an infinite supply of \emph{individual
		variables} $x,x_1,x_2,\dots$ and of \emph{set variables}
	$X,X_1,X_2,\dots$. The \emph{atomic formulas} are 
	$V x$ (``$x$ is a vertex''), $E y$ (``$y$ is an edge''), $I xy$ (``vertex $x$
	is incident with edge $y$''), $x=y$ (equality),
	$P_a x$ (``vertex or edge $x$ has label $a$''), and $X x$ (``vertex or
	edge $x$ is an element of set $X$'').  \emph{MSO formulas} are built up
	from atomic formulas using the usual Boolean connectives
	$(\lnot,\land,\lor,\rightarrow,\leftrightarrow)$, quantification over
	individual variables ($\forall x$, $\exists x$), and quantification over
	set variables ($\forall X$, $\exists X$).

	\emph{Free and bound variables} of a formula are defined in the usual way. To indicate that the set of free individual variables of formula $\Phi$
	is $\{x_1, \dots, x_\ell\}$ and the set of free set variables of formula $\Phi$
	is $\{X_1, \dots, X_q\}$ we write $\Phi(x_1,\ldots, x_\ell, X_1,
	\dots, X_q)$. If $G$ is a graph, $v_1,\ldots, v_\ell\in V(G)\cup E(G)$ and $S_1, \dots, S_q
	\subseteq V(G)\cup E(G)$ we write $G \models \Phi(v_1,\ldots, v_\ell, S_1, \dots, S_q)$ to denote that
	$\Phi$ holds in $G$ if the variables $x_i$ are interpreted by the vertices or edges $v_i$, for $i\in [\ell]$, and the variables $X_i$ are interpreted by the sets
	$S_i$, for $i \in [q]$.

	The following result (the well-known Courcelle's theorem~\cite{Courcelle90}) 
	shows that if $G$ has bounded treewidth~\cite{RobertsonS84} then we
	can find an assignment $\varphi$ to the set of free variables $\mathcal{F}$ with $G \models \Phi(\varphi(\mathcal{F}))$ (if one exists) in linear time. 

% \todo{When using statements from literature, we are currently using ``Proposition'' in other places. Change this ``fact'' to ``Proposition''?- RMS}

	\begin{longproposition}[Courcelle's Theorem~\cite{Courcelle90,ArnborgLS91}]\label{fact:MSO} 
Let $\Phi(x_1,\dots,x_\ell, X_1,\dots, X_q)$ be a fixed MSO formula with free individual variables $x_1,\dots,x_\ell$ and free set variables $X_1,\dots,X_q$, and let $w$ be a
		constant. Then there is a linear-time algorithm that, given a labeled
		graph $G$ of treewidth at most $w$, either outputs  $v_1,\ldots, v_\ell\in V(G)\cup E(G)$ and $S_1, \dots, S_q	\subseteq V(G)\cup E(G)$ such that $G \models \Phi(v_1,\ldots, v_\ell, S_1, \dots, S_q)$ or correctly identifies that no such vertices $v_1,\ldots, v_\ell$ and sets $S_1, \dots, S_q$ exist.
	\end{longproposition}
	\fi

\subparagraph{Rooted graphs.} A rooted graph is a graph where some vertices are labeled---formally:
%. A formal definition is as follows. 

\begin{definition}[Rooted graphs]\label{def:boundariedGraph}
A {\em rooted graph} is a graph $G$ with a set $\partial(G)\subseteq V(G)$ of distinguished vertices
 called {\em roots} and a labeling $\lambda_G: \partial(G)\rightarrow \mathbb{N}$. 
The set $\partial(G)$ is the {\em root set} of $G$, and the {\em label set} of $G$ is $\Lambda(G)=\{\lambda_{G}(v)\mid v\in \partial(G)\}$.
\end{definition}

\begin{definition}[{Topological minors of rooted graphs~\cite{GroheKMW11}}]
Let $G$ and $H$ be two rooted graphs with labelings $\lambda_H$ and $\lambda_G$ for their root sets.  We say that $H$ is a {\em topological minor} of $G$ if there exist injective functions $\phi: V(H)\rightarrow V(G)$ and $\varphi: E(H)\rightarrow {\sf paths}(G)$ such that 
\begin{itemize}
\setlength\itemsep{0em}
\item for every $e=\{h,h'\}\in E(H)$, the endpoints of $\varphi(e)$ are $\phi(h)$ and $\phi(h')$,
\item for every distinct $e,e'\in E(H)$, the paths $\varphi(e)$ and $\varphi(e')$ are internally vertex-disjoint,
\item there does not exist a vertex $v$ in the image of $\phi$ and an edge $e\in E(H)$ such that $v$ is an internal vertex on $\varphi(e)$, and
\item for every $v\in \partial(H)$, $\lambda_H(v) = \lambda_G(\phi(v))$.
\end{itemize}
We say that $(\phi,\varphi)$ is a {\em realization} of $H$ as a topological minor in $G$. Moreover, the subgraph of $G$ induced by the union of the edges in the paths in $\varphi(e)$ for $e\in E(H)$ and the vertices in $\phi(h)$ for $h\in V(H)$ is called a {\em topological minor model} of $H$. 
\end{definition}
Note that if $\partial(G)=\emptyset$, then the above coincides with the usual definition of topological minors.  
%the definition above coincides with the standard definition of a topological minor.  

\begin{proposition}{\rm (\cite{GroheKMW11})}\label{prop:TMcontainment}
	There is
	 an algorithm that, given two rooted graphs $G$ and $H$, 
     % with labelings $\lambda_H$ and $\lambda_G$ respectively,  
     runs in time $f(|H|)\cdot \bigoh(|G|^{3})$ for some 
	 computable function $f$ and decides whether $H$ is a topological minor of $
     G$.\end{proposition}
  % \todo{I: Maybe give more details to self-reducibility? Added - RS}

\begin{remark}\label{rem:self-reduceTMcontainment}
 
 \iflong
Although the algorithm in Proposition~\ref{prop:TMcontainment} is for the decision version of the problem, a realization if one exists, can be computed using self-reducibility arguments as  follows. By iteratively deleting an edge of $G$ and running the algorithm of Proposition \ref{prop:TMcontainment}, we can identify a minimal subgraph $G'$ of $G$ that is a subdivision of $H$. Isolated vertices of $H$ can be mapped to isolated vertices of $G'$ in an arbitrary way. The number of vertices of $G'$ of degree at least three is bounded by $|V(H)|$, and the mapping of these to $V(H)$ can be guessed.  Moreover, the number of maximal paths in $G'$ that only have degree-2 vertices as internal vertices (call this set $\cal P$) must be bounded by a function of $|V(H)|$, and we can guess the mapping of vertices of $H$ of degree at most 2 into $\cal P$ and the final verification step is straightforward brute force. 
\fi
\end{remark}

\newcommand{\td}[0]{\mathsf{td}}
\iflong
\begin{longdefinition}[Forest embedding and treedepth]\label{def:tree-depth}{\rm 
A \emph{forest embedding} of a graph $G$ is a pair $(F,f)$, where $F$ is a rooted forest and $f : V(G) \rightarrow V(F)$ is a bijective function, such that for each $\{u,v\} \in E(G)$, either $f(u)$ is a descendant of $f(v)$, or $f(v)$ is a descendant of $f(u)$. The \emph{depth} of the forest embedding is {the number of vertices in the longest root-to-leaf path in $F$}. The \emph{treedepth} of a graph $G$, denoted by $\td(G)$, is the minimum over the depths of all possible forest embeddings of $G$. 
}
\end{longdefinition}

If $G$ is connected, then the forest embedding of $G$ consists of a single tree, i.e., it is a tree embedding. 

 \begin{longproposition}[\cite{sparsity}]\label{proptreedepth}
If a graph $G$ has no path of length at least ${\ell}$, then it has treedepth at most~$\ell$. 
 \end{longproposition}

\begin{longobservation}
    \label{separatorsInTreedepth}
    Let $G$ be a graph with a forest embedding $(F,f)$ and consider some vertex $v$. If the number of children of $v$ in $F$ is at least $\alpha$ and $Z$ denotes the set of vertices appearing in the root-to-$v$ path in the tree of $F$ containing $v$, then the number of connected components of  $G-Z$ is at least $\alpha$. 
\end{longobservation}
\begin{proof}
By the definition of forest embeddings,  (i) every descendant of $v$ in $F$ is adjacent in $G$ to either another descendant of $v$ in $F$ or a vertex in $Z$ and (ii) descendants of distinct children of $v$ in $F$ are non-adjacent in $G$. 
\end{proof} \fi

\subparagraph*{Problem Formalizations.}
In our problems of interest, we are given an undirected graph $G$ and a set $\R=\{R_1, R_2, \ldots, R_k\}$ of $k$ robots where $\R$ is partitioned into two sets $\M$ and $\F$. Each $R_i \in \M$, has a starting vertex $s_i$ and a destination vertex $t_i$ in $V(G)$ and each $R_i \in \F$ is associated only with a starting vertex $s_i \in V(G)$.
We refer to the elements in the set $\{s_i\mid i\in [k]\}\cup \{t_i\mid R_i\in \M\}$ as {\em terminals}. 
%The set $\M$ contains robots that have specific destinations they must reach, whereas $\F$ is the set of remaining ``free'' robots. 
We assume that all the $s_i$ are pairwise distinct and that all the $t_i$ are pairwise distinct.  We use a discrete time frame $[0, q]$, $q \in \nat$,
% \todo{I: You use $q$ here and later use $t$. RMS: Fixed.} 
to reference the sequence of moves of the robots. In each time step $x \in [0, q]$, exactly one robot moves and the rest remain stationary. 

In \textsc{Coordinated Sliding-Motion Planning} (CSMP), we are given a tuple $(G, \R=(\M,\F), k, \ell)$, 
where $G, \R$ and $k$ are as described in the last paragraph and $\ell\in {\mathbb N}$. 
% graph, $\R=\{R_i \mid i \in [k]\}$ is a 
% set of robots partitioned into sets $\cal M$ and $\cal F$, where each robot in $\cal M$ is given as a pair 
% of vertices $(s_i, t_i)$  and each robot in $\cal F$ as a single vertex $s_i$, and $k, \ell \in \nat$.
The goal is to decide whether there is a sequence of $q\leq \ell$ moves such that robots move serially one at a time, where a robot's move consists of a move along a free/unobstructed path (i.e., a path in which no other robot is located at the time of the move), and such that each $R_i\in \R$ starts in $s_i$ at time step $0$ and each $R_i \in \M$ is positioned on $t_i$ at the end of time step $q$. 
% In {\planaroneagentproblem}, we additionally have that $G$ is a planar graph and $\M$ is a singleton.
The second problem that we consider is a restriction of 
\cmpl{} to the case where $G$ is planar and $\M$ contains exactly one robot, referred to as the \emph{main} robot. We refer to this restriction as \planaroneagentproblem, and  for simplicity represent its instances as tuples of the form $(G,s,t,\ell,\R)$, where $s$ and $t$ are the starting and destination vertices of the main robot, respectively, and $\R$ is the set of all robots.

% \todo{I: Should we define $G_i$ (or a labeled graph) to be the graph and robots on it at step $i$? RMS: this is not used in my section. \\I: Maybe we should keep your definition here but give a two-sentence informal definition before it, and say that now we define the problem formally as this will be needed for the technical parts of the paper.}

We next introduce some notation that will be used in our technical sections. 
 A \emph{route} for robot $R_i$ is a tuple $W_i=(u^{i}_0,P^{i}_1,u^{i}_1, \ldots, P^{i}_{q},u^{i}_{q})$
 % \todo{I: $t$ should be avoided everywhere since it is used as the destination of the main robot. RMS: changed}
 where each $u^{i}_j$ is a vertex in $G$ and each $P^{i}_j$ is a simple path in $G$,   such that (i) $u^{i}_0=s_i$ and $u^{i}_{q}=t_i$ if $R_i \in \M$ and (ii) $\forall j \in [q]$, $P^{i}_j$ is a $u^{i}_{j-1}$-$u^{i}_{j}$ path in $G$ (if $u^{i}_{j-1}=u_j^{i}$, then $P^{i}_j$ is the singleton path comprising the same vertex).
  
 Put simply, each $W_i$ corresponds to a ``walk'' in $G$.  If vertices of $W_i$ at two consecutive time steps $j-1$ and $j$ are identical, then $j$
 % it\todo{I: What is ``it''? RMS: fixed.} represents 
is  a {\em waiting time step} for the robot $R_i$. 
 Moreover, each $R_i$ begins at its starting vertex at time step $0$, and if $R_i \in \M$ then $R_i$ is at its destination vertex at time step $q$. 
 Though we work with undirected graphs and the paths described above are undirected paths, each path $P^{i}_j$ has a natural {\em starting vertex} and {\em ending vertex} designated by the time step $j$. 
 
 \begin{definition} \rm Two routes $W_i=(u^{i}_0,P^{i}_1, \ldots, P^{i}_q,u^{i}_{q})$ and $W_j=(u^{j}_0,P^{j}_1, \ldots, P^{j}_q,u^{j}_{q})$, where $i \neq j \in [k]$, are \emph{non-conflicting} if $\forall r \in \{0, \ldots, q\}$, 
 % $u^{i}_r \neq u^{j}_r$, and (ii) 
 the path $P^{i}_r$ is disjoint from $u^{j}_{r-1}$ and the path $P^{j}_r$ is disjoint from $u^{i}_{r-1}$. 
Otherwise, we say that $W_i$ and $W_j$ \emph{conflict}. 
\end{definition}

In particular, in the above definition, for every $i \neq j \in [k]$ and  $\forall r \in \{0, \ldots, q\}$, $u^{i}_r \neq u^{j}_r$. 
\iflong Intuitively, two routes conflict if the corresponding robots are at the same vertex at the end of a time step, or one ``hits'' the other while traveling along a path during some  time step. \fi
A \emph{schedule} $\SSS$ for $\R$ is a set of pairwise non-conflicting routes $W_i, i \in [k]$, during a time interval $[0, q]$, where exactly one robot moves in each time step.  The {\em number of moves} in a route $W_i=(u^{i}_0,P^{i}_1, \ldots, P^{i}_q,u^{i}_{q})$ or the number of moves of its associated robot is the number of time steps $j$ such that $u^{i}_j\neq u^{i}_{j+1}$. The \emph{total number of moves} (or \emph{makespan}) in a schedule is the sum of the number of moves in its routes. 
A schedule of minimum total moves is called an {\em optimal schedule}.

% Using the introduced terminology, we formalize the \textsc{Coordinated Sliding-Motion Planning} (CSMP). 
% The input is a tuple $(G, \R=(\M,\F), k, \ell)$, where $G$ is a 
% graph, $\R=\{R_i \mid i \in [k]\}$ is a 
% set of robots partitioned into sets $\cal M$ and $\cal F$, where each robot in $\cal M$ is given as a pair 
% of vertices $(s_i, t_i)$  and each robot in $\cal F$ as a single vertex $s_i$, and $k, \ell \in \nat$.
% The goal is to decide whether there is a schedule for $\R$ with at most $\ell$ moves. 

% A schedule of minimum total moves is called an {\em optimal schedule}. 
% 
% 
% 
% 
% I
% The second problem that we consider is a restriction of the CSMP to the case where $G$ is planar and $\M$ contains exactly one robot, referred to as the \emph{main} robot. That is, there is exactly one robot with a destination. We refer to this restriction as \planaroneagentproblem.
% 
A vertex $v$ is \emph{occupied} at time step $i$ if some robot is located at $v$ at time step $i$; otherwise, $v$ is said to be \emph{free} (or \emph{unoccupied}). If the time step is not specified, it is assumed to be time step $0$ (i.e., in the initial state). At each time step, a robot may either move to a different vertex, or stay at its current vertex. 
Finally, we always assume that the input graph is connected since otherwise, we can work on individual connected components.
% A robot's move is a slide along an unobstructed path in $G$.

% For a graph $G$, ${\sf paths}(G)$ denotes the set of all simple paths in $G$.
% The length of a path is the number of edges in it.
%  For a path $P$ and vertices $u,v$ in it, we denote by $P[u,v]$ the subpath of $P$ starting at $u$ and ending at $v$.  The \emph{$P$-distance} between $u$ and $v$, denoted $\dist_{P}(u, v)$, is the length of $P[u, v]$. We write $[n]$, where $n \in \nat$, for $\{1, \ldots, n\}$.

\subparagraph*{Remarks on Feasibility and Constructiveness.} When only considering feasibility, \cmpl{} is equivalent to the version where each move is along a single edge instead of an unobstructed path. 
 
\iflong
Hence, by the result of Yu and Rus~\cite{YuR14} (observed to be extendable to the setting where not every robot has a destination \cite{DeligkasEGK024}), it follows that it is linear-time solvable. 

\begin{longproposition}[Implied by \cite{YuR14}]\label{prop:reconfiguration}
    The existence of a schedule for an instance of \cmpl{} can be decided in linear time. 
    Moreover, if such a schedule exists, then a schedule with  $\Oh(|V(G)|^3)$ moves can be computed in $\Oh(|V(G)|^3)$ time. 
\end{longproposition}

 Proposition~\ref{prop:reconfiguration} implies inclusion in \NP, and allows us to assume henceforth that every instance of {\cmpl} is feasible (otherwise, in linear time we can reject the instance). 
  \fi
  
 We note that even though {\cmpl} is formalized as a decision problem, all the algorithms provided in this paper are constructive and can output a corresponding schedule (when it exists) as a witness.

\section{FPT Algorithm for \cmpl{} Parameterized by the Number of Robots}

In this section, we show that \cmpl{} is FPT parameterized by the number $k$ of robots in the input. Note that we do not assume any restrictions on the input graph.

Our strategy has three high-level steps: (i) We first show that if a solution exists, then there is an optimal schedule whose makespan is bounded\footnote{We recall that by bounded, we mean ``bounded by a function of the parameter''---in this case, $k$.}. (ii) We then show that such a solution is essentially a realization of a rooted graph $H$ of bounded size as a topological minor in a rooted graph $G'$ obtained by assigning unique labels to the terminals in $G$. We also prove the converse, that is, from any realization of $H$ as a topological minor in $G'$, one can obtain an optimal solution for the given instance.  (iii) Finally, we leverage known results on topological minor containment to verify the existence of the correct rooted graph. Let us now give more detail on Steps (i) and (ii). 

\noindent
Step (i): To prove the existence of a solution with boundedly many moves, we preprocess the graph by applying a reduction rule that shortens degree-2 paths to a bounded length. This simplification allows us to focus on two cases for the graph (assuming it is connected):

 Case (a): Bounded Treedepth.   If the graph has bounded treedepth, we show that instances with bounded treedepth and a bounded number of robots can be reduced iteratively without altering the solution. This reduction process results in an instance of bounded size, implying the existence of a solution with a bounded makespan.

 Case (b): Long Paths. Suppose every vertex is the endpoint of a sufficiently long path. We use ideas from \cite{DeligkasEGK024} to define a suitable notion of {\em havens} for our setting; in particular, havens are subgraphs of bounded size that can efficiently handle collisions. Here, by the ``efficient handling'' of collisions, we mean that given any pair of ``configurations'' of robots in a haven, we can move the robots from one configuration to another without any conflicts using only boundedly many moves.
  % instance of {\cmpl} where the input graph is a haven
 % \todo{I: Redescribe in terms of subgraphs} 
 % has a solution 
 % with
  Thus, with a small number of time steps, a robot can pass through a haven without collision, regardless of other robots in the haven.
 % with a small number of ``resolving steps'' between the time step it enters the haven and the time step it exits,  we can enable it to pass through the haven without a collision. 
 Building on this fact, we construct a schedule where each robot interacts with only a bounded number of havens. Within each haven, the number of moves used by a robot is bounded and we show that outside the havens, the robot only makes boundedly many moves as well. This yields a bound on the makespan of the schedule.

\noindent
Step (ii): Consider the graph induced by the edges in the solution. We partition the vertices of this graph into two sets: important and unimportant vertices. The important vertices are those with degree at least three, the terminals, and any vertex where a robot waits at any time step.
We show that the number of important vertices is bounded by a function of $k$ (say, $\zeta(k)$). Moreover, by definition, the unimportant vertices form paths with only degree-2 vertices as internal vertices. Additionally, every robot that visits a vertex in such a path slides without stopping on any vertex of this path. This structure enables us to show that the graph induced by the edges in the solution (with terminals as roots) is a realization of some rooted graph with at most  $\zeta(k)$ vertices as a topological minor in the input graph.

\subsection{Bounding the Makespan in an Optimal Schedule}

Our first task is to apply a reduction rule to the given instance that enforces useful structural properties on the graph without affecting the existence of a solution.

\newtheorem{reduction}{Reduction Rule}
\iflong
\begin{reduction}
\label{shortcircuit}
	Let $I=(G, \R=(\M,\F), k, \ell)$ be an instance of {\cmpl}. If there is a path $P$ in $G$ such that every internal vertex of $P$ is disjoint from the set of terminals and has degree exactly two in $G$, then shorten $P$ to length $min\{|P|,2k+1\}$. 
\end{reduction}

\begin{longlemma}\label{lem:correctnessShortcircuit}
	Let $I=(G, \R=(\M,\F), k, \ell)$ be an instance of {\cmpl} and $I'$ be the instance obtained by applying Reduction Rule \ref{shortcircuit}. These two instances are equivalent, that is, $I$ has a solution if and only if $I'$ has a solution.	
\end{longlemma}
\fi
\iflong \begin{proof}
Consider a maximal path $P$ in $G$ such that every internal vertex has degree exactly two in $G$ and has no terminal as an internal vertex. Suppose that $P$ has more than $2k$ internal vertices and let $x$ and $y$ be its endpoints. Let $P'$ denote the shortened $x$-$y$ path with which we replaced $P$. Let $I'$ denote the instance resulting from applying the reduction rule. 
For each endpoint $z\in \{x,y\}$ of $P'$, let $L_{z}$ denote the $k$ internal vertices of $P'$ closest to $z$. It is straightforward to see that if $I'$ has a solution, then so does $I$. Conversely, we consider a schedule $\SSS=\{W_i\mid  i \in [k]\}$ for $I$ and produce a schedule $\SSS'$ for $I'$ with at most the same number of moves as $\SSS$. The idea is to ensure that at any time step, (i) the robots waiting on the path $P$ are the same as those waiting on $P'$, and (ii)  the relative ordering of the robots waiting on the path $P$ is the same as their relative ordering on the path $P'$. Here, by the relative ordering of the robots on a path, we refer to the permutation of robots that gives their ordering in terms of increasing distance along $P$ from the endpoint whose identifier is lexicographically the smaller.  This is ensured by considering the following four cases and constructing $\SSS'$ accordingly. 

\begin{itemize}
\item If there is a time step $j$ in the schedule $\SSS$ during which a robot $R_i$ enters $P$ through one of the endpoints of $P$ and exits through the other, then the same movement is captured in $\SSS'$ by passing through the entirety of $P'$. 
\item Next, consider a time step $j$ where robot $R_i$ enters $P$ through the endpoint $z\in \{x,y\}$  
and waits on an internal vertex of $P$. Recall that $u^{i}_j$ is the position of robot $R_i$ at the end of time step $j$. Then, formally, we have that $u^{i}_{j-1}$
% \todo{I: Maybe remind the reader what these vertices are.}  
is not in $P$, $u^{i}_{j}$ is an internal vertex of $P$ and the path $P^{i}_j$ contains $z$. Then, in time step $j$ of $\SSS'$, we let $R_i$ enter $P'$ through $z$ and make it wait at the vertex of $L_z$ that is farthest from $z$ among all vertices of $L_z$ that do not have a robot waiting on them.  Since $L_z$ has size $k$ and by our strategy, robots occupy $L_z$ starting from the farthest available vertex from $z$, it follows that every robot that needs to wait on $P'$ in this way always has an available vertex.
\item Next, consider a time step $j$ where robot $R_i$ exits the path $P$ through an endpoint $z\in \{x,y\}$ after waiting on an internal vertex at time step $j-1$. Formally,  $u^{i}_{j-1}$ is in $P$, $u^{i}_{j}$ is not in $P$ and the path $P^{i}_j$ contains $z$. Then, we move robot $R_i$ to $z$ and then follow the rest of the path as given by $\SSS$. 
\item Finally, consider the case where a robot $R_i$ moves from one internal vertex of $P$ to another. In this case, we do not need to make a move in $\SSS'$ since we trivially maintain both the subset of robots on $P'$ and their relative ordering. 
\end{itemize}

In every case, assuming that the relative ordering of the robots on $P$ and $P'$ is the same at time step $j-1$, it follows that the specified moves are possible and moreover, we  ensure that the relative ordering of the robots on the paths $P$ and $P'$ is the same at the end of time step $j$ (in particular, the subset of robots on both paths is the same). Moreover, by construction of $\SSS'$, we have also ensured that a vertex $v$ disjoint from $P$ has robot $R_i$ on it at time step $j$ in schedule $\SSS$ if and only if the same happens in schedule $\SSS'$. This ensures that $\SSS'$ has at most the same number of moves as $\SSS$ and no pair of routes in $\SSS'$ conflict,  completing the proof. 
\end{proof} \fi

We say that an instance $I$ is {\em irreducible} if Reduction Rule \ref{shortcircuit} cannot be applied. In the rest of this section, we work with such instances. Our goal now is to show that every \YES-instance has a schedule in which the makespan is bounded by a function of the number of robots. \iflong Specifically, we prove the following fact.\fi 

\iflong
\begin{restatable}{lemma}{stepbound}

   \label{lem:stepbound}
Let $I=(G, \R=(\M,\F), k, \ell)$ be an instance of {\cmpl}. If $I$ is a \YES-instance, then $I$ has an optimal schedule with  $2^{\bigoh(k^{4})}$ moves.
  
\end{restatable}
\fi 

The proof strategy of Lemma \ref{lem:stepbound} is as follows. We show that for a \YES-instance, either the treedepth of the input graph is bounded or a certain locality property holds (roughly speaking, every vertex is sufficiently close to a vertex of degree at least three). We then show that in either case, there is a solution with boundedly-many moves. Specifically, when the treedepth is bounded, we show how to reduce the graph to a bounded-size graph without affecting the existence of a solution (thus implicitly bounding the makespan of an optimal solution) and otherwise, we show how the aforementioned locality property can be used along with a greedy strategy to produce a schedule with boundedly many moves (again, bounding the makespan of an optimal solution).

\iflong
Towards the proof of Lemma \ref{lem:stepbound}, we adapt ideas from \cite{DeligkasEGK024} on the notion of ``havens'' that was originally designed to address conflicts occurring in schedules for \cmpl{} instances where every move is required to be along a single edge.

\begin{longdefinition} 
% \rm [Nice Vertex]
\label{def:nicevertex}
For $q\in {\mathbb N}$ and graph $G$, a vertex $w\in V(G)$ is a $q$-\emph{anchor} if it has degree at least three in $G$ and lies on a path $P$ of length $q$. This path $P$ is called a $q$-haven anchored at $w$. We also say that {\em $P$ is a $q$-haven for} a vertex $v\in V(G)$ if $v$ lies on $P$ and has $P$-distance at least $\lceil q/3\rceil $ and at most $\lfloor 2q/3\rfloor$ from the anchor $w$. Additionally, we say that $P$ is a {\em strong} $q$-haven anchored at $w$ if it is a $q$-haven anchored at $w$ for both endpoints of $P$.
\end{longdefinition}

Roughly speaking, a strong $q$-haven for a vertex $v$ anchored at a vertex $w$ of degree at least three is a path of length $q$ starting at $v$ and containing $w$ such that $w$ is approximately in the middle of the path. If $q$ depends only on $k$, then such a path can be viewed as being short enough to allow local modifications in the schedule within the path at a relatively low cost. At the same time, since $w$ is sufficiently far from either endpoint, it allows ample room on either side of $w$ to place all the robots if necessary. This property will prove to be useful in subsequent arguments.

\begin{longlemma}\label{lem:nearHaven}
	In an irreducible instance $I=(G, \R=(\M,\F), k, \ell)$, either the treedepth of $G$ is  $\bigoh(k^{2})$ or, for every vertex $v$, there is a path $H$ and a vertex $w$ such that (i) $v$ is one of the endpoints of $H$ and (ii) $H$ is a strong $q$-haven anchored at $w$, where $q=\Theta(k^{2})$.
\end{longlemma}

\begin{proof}
Fix some $\alpha=\Theta(k^{2})$ with the precise value to be determined later. 
If $G$ has no path of length $\alpha$, then by Proposition \ref{proptreedepth}, the treedepth of $G$ is bounded by $\alpha$.

Otherwise, consider a path $P$ of length $\alpha$ and pick a vertex $v\in V(G)$. Let $P'$ be a path from $v$ to an arbitrary vertex in $P$. Notice that the union of $P$ and $P'$ contains a $v$-$u$ path $P''$ of length exactly $\alpha/2$ 
% (we later pick $\alpha$ to be some multiple of 12)
for some vertex $u$.
Divide $P''$ into three equally-long subpaths starting from $v$: $P_1''$, $P_2''$, $P_3''$. Due to the fact that there are at most $2k$ terminals, it follows that $P_2''$ either contains a vertex with degree at least three in $G$, or it contains at least one subpath of length at least $\alpha/(3\cdot 2\cdot 2k)$ that has only degree-2 vertices as internal vertices. However, by irreducibility, we have that $\alpha/12k\leq 2k$. 
By choosing $\alpha=24k^{2}$, we infer that $P_2''$ contains a vertex of degree 3, which we set as the required vertex $w$ and the path $P''$ as the path $H$ claimed in the lemma statement. 
\end{proof}

The following lemma describes the reduction procedure that ultimately enables us to reduce any sufficiently large graph of low treedepth without affecting a solution.

\begin{longlemma}\label{lem:pruningLemma1}
Consider an instance $I=(G, \R=(\M,\F), k, \ell)$ and a vertex set $X$ in $G$ of size at most $d$. 
     If the number of connected components of $G-X$ is greater than $3k\cdot 2^{d}$, then there is a connected component $C$ in $G-X$ such that $I$ is a \YES-instance if and only if $I'=(G-C, \R=(\M,\F), k, \ell)$ is a \YES-instance. 
     \end{longlemma}

\begin{proof}
The proof idea is as follows. If the premise of the lemma is satisfied, then there exists a set ${\cal C}=C_1,\dots, C_{3k+1}$ of connected components  of $G-X$ such that $N(C_i)=N(C_j)$ for every $i\neq j$. At most $2k$ of the components in $\cal C$ contain terminals. Hence, we may assume without loss of generality that the components in $\CCC'=\{C_1,\dots,C_{k+1}\}$ do not contain terminals. Call these components, {\em exceptional components}, and the vertices in them {\em exceptional vertices}.

We then argue that removing the component $C_{k+1}$ does not affect the existence of a solution. To do so, we show that an optimal schedule for $I$ can be modified in such a way that if robot $R_i$ interacts with the exceptional vertices (so, in particular, the component $C_{k+1}$), it is instead re-routed through the component $C_i$ and stays disjoint from the remaining exceptional components. In this way, by ensuring that $C_i$ is `reserved' for the exclusive use of robot $R_i$, we argue that there are no conflicts caused by our re-routing. The formal argument follows. 

Let $I'=(G-C_{k+1}, \R=(\M,\F), k, \ell)$. It is clear that if $I'$ is a \YES-instance, then so is $I$. So, let us consider the converse and take an optimal schedule $\SSS=\{W_i\mid i\in [k]\}$ for $I$. Our goal is to produce a schedule $\SSS'$ for $I'$ with no more moves than in $\SSS$. For robots that do not visit an exceptional vertex, the route in $\SSS'$ is set to be the same as that in $\SSS$. 
If no exceptional component is ever visited by any robot in the schedule $\SSS$, then we are done.
So, assume that this is not the case. For every $i\in [k$], let $j_i$ be the first time step during which robot $R_i$ visits an exceptional component 
and let $j'_i$ be the last time step during which robot $R_i$ visits an exceptional component.
Notice that $j_i$ and $j'_i$ need not exist for every robot $R_i$. However, if $j_i$ exists for a robot $R_i$ that has a destination, then $j'_i$ also exists since no terminals are present in any component in $\CCC'$.

We now describe how route $W_i$ is modified for a robot $R_i$ for which time step  $j_i$ exists. 
Let the path $P^{i}_{j_i}$ be an $x$-$y$ path for vertices $x$ and $y$. We consider the following two scenarios based on the first interaction of robot $R_i$ with the exceptional components. By the definition of $j_i$, it must be the case that $x$ is not an exceptional vertex.

\begin{enumerate}
    
    \item Suppose that $y$ is not an exceptional vertex. That is, in time step $j_i$, robot $R_i$ simply `passes through' the exceptional vertices without waiting on any of them.
    Let $x'$ and $y'$ be vertices of the set $X$ that appear on this path right before the first occurrence of an exceptional vertex and right after the last occurrence of an exceptional vertex, respectively. We then modify $P^{i}_{j_i}$ by removing the $x'$-$y'$ subpath and replacing it with an $x'$-$y'$ path that has all internal vertices in $C_i$.  The required $x'$-$y'$ path exists since every exceptional component has the same neighborhood. Moreover, as long as we ensure that no other robot visits a vertex of $C_i$, this does not create a conflict.  
    
    \item Suppose that $y$ is an exceptional vertex. That is, at the end of time step $j_i$, robot $R_i$ waits at an exceptional vertex.   Let $x'$ be the vertex of $X$ that appears on $P^{i}_{j_i}$ right before the first occurrence of an exceptional vertex. Then, we replace the $x'$-$y$ subpath with an $x'$-$y'$ edge where $y'$ is an arbitrary neighbor of $x'$ in $C_i$. We then consider two subcases:

    \begin{enumerate}        
        \item If $j_i'$ does not exist, it must be the case that $R_i$ does not have a destination and so, we never move $R_i$ again.  
        \item Otherwise, consider the path $P^{i}_{j_i'}$ using which robot $R_i$ leaves the exceptional components for the final time. Let $\hat y$ be the last occurrence of an exceptional vertex in this path. Then, in $P^{i}_{j_i'}$, replace the subpath from the starting vertex up to $\hat y$ with an arbitrary path from $y'$ to $\hat y$ that has all vertices except $\hat y$ inside $C_i$, which must exist since all exceptional components have the same neighborhood. 
    \end{enumerate}

    As long as we ensure that in $\SSS'$, no other robot visits a vertex of $C_i$, these operations do not create a conflict.
\end{enumerate}

It follows from the construction that if $j_i$ does not exist, then the route of the robot $R_i$ will be the same in $\SSS$ and $\SSS'$.
More generally, at any time step $j$, robot $R_i$ and vertex $v$ that is not exceptional, robot $R_i$ is on vertex $v$ at the end of time step $j$ in $\SSS$ if and only if the same happens in $\SSS'$. This also implies that the number of moves in $\SSS'$ is no more than the number of moves in $\SSS$. Finally, since our construction ensures that in $\SSS'$, robot $R_i$ only visits components $C_i$ among the exceptional components, we do not create conflicts by our re-routing. 
\end{proof}

\begin{longlemma}\label{lem:mainTreedepth}
% Suppose that $G$ has treedepth bounded by $\bigoh(k^{2})$.
If $I$ is a \YES-instance, then it has a schedule with at most $k^{3}\cdot 2^{\bigoh(\td(G)^{2})}$ steps.
% for a computable function $f$.
\end{longlemma}

\begin{proof}
 We begin by exhaustively enumerating every set $X$ in $G$ of size at most $\td(G)$. For those sets $X$ of size  $d\leq  \td(G)$ such that the number of connected components of $G-X$ is greater than $3k\cdot 2^{d}$, remove the `irrelevant' component given by Lemma \ref{lem:pruningLemma1}. Note that we do not care about the algorithmic efficiency of this operation as the lemma statement asserts the existence of a certain schedule and not the efficient computability of this schedule.
 Moreover, removing a vertex does not increase the treedepth of a graph and so, we conclude that when this reduction is no longer applicable, the graph, $G'$, also has treedepth at most $\td(G)$. Finally, recall that we work on a connected graph $G$ and notice that $G'$ is also connected and by Lemma \ref{lem:pruningLemma1}, since $I$ is a \YES-instance, so is the instance induced by $G'$. 

We claim that the number of vertices in $G'$ is bounded by $3k\cdot 2^{\td(G)^{2}}$. Consider the tree embedding of $G'$. If every vertex has at most $3k\cdot 2^{\td(G)}$ children, then we are done since the depth of the tree embedding is at most $\td(G)$. Suppose that there is a vertex $v$ that has more than $3k\cdot 2^{\td(G)}$ children in the forest embedding. 
 Let $Z$ be the set of vertices on the root-to-$v$ path in the embedding. Notice that $Z$ has size at most $d$. Let $\cal C$ denote all the connected components of $G'-Z$. By Observation \ref{separatorsInTreedepth}, 
 it follows that $|\cal C|$ has size more than $3k\cdot 2^{\td(G)}$, which is a contradiction to the fact that $G'$ cannot be reduced further. 

Finally, we use the fact that every feasible instance on a graph with $n$ vertices has a schedule with $\bigoh(n^{3})$ steps (Proposition \ref{prop:reconfiguration}) to conclude that $G'$ has a schedule with $k^{3}\cdot 2^{\bigoh(\td(G)^{2})}$ steps. 
\end{proof}

% \begin{remark}
%     Lemma \ref{lem:mainTreedepth} can be make constructive and implies an FPT algorithm for \cmpl{} parameterized by $\td$ and number of robots. We later prove that \cmpl{} is FPT parameterized by the number of robots alone. 
% \todo{Is the problem FPT by treedepth alone? If yes, is it easy to see?}
% \end{remark}

Since Lemma \ref{lem:mainTreedepth} achieves our objective of bounding the size of the optimal schedule when the graph has bounded treedepth, we next focus on the case where Lemma \ref{lem:nearHaven} is only able to guarantee that for every vertex $v$, there is a $q$-haven for $v$ anchored at some vertex $w$, where $q=\Theta(k^{2})$.

For a set $S \subseteq \R$ of robots and a subgraph $H$, a \emph{configuration} of $S$ w.r.t.~$H$ is an injection $\iota: \  S \longrightarrow V(H)$. We show using a lemma of Deligkas et al. (Lemma 10, \cite{DeligkasEGK024}, see arXiv version~\cite{DeligkasEGK024Arxiv} for the proof) that we can move the robots in a $\Theta(k^{2})$-haven from any configuration to any other configuration within the same haven, using a total of $\Oh(k^6)$ moves. 

\begin{longlemma}[Slightly modified statement of Lemma 10, \cite{DeligkasEGK024}]\label{lem:oldLemSwap}
Consider a vertex $w$ and three connected subgraphs $C_1, C_2, C_3$ of $G$ such that: (i) the pairwise intersection of the vertex sets of any pair of these subgraphs is exactly $w$, and (ii) $|V(C_1)| \geq k+1$, $|V(C_2)| \geq k+1$, and $|V(C_3)| \geq 2$. Let $H$ denote the subgraph  of $G$ induced by the vertices in $\{x\} \cup V(C_1) \cup V(C_2)$ whose distance from $w$ in $H$  is at most $p$. Then for any set of robots $S \subseteq \R$ in $H$ with current configuration $\iota(S)$, any configuration $\iota'(S)$ with respect to $H$ can be obtained from $\iota(S)$ via a sequence of $\Oh(k^{2}\cdot p)$ moves that take place in $H$.
\end{longlemma}

\begin{remark*}
Note that in Lemma \ref{lem:oldLemSwap} as stated in \cite{DeligkasEGK024}, each move is along a single edge. However, in this paper we allow moves along unobstructed paths, so the number of moves given by their lemma is an upper bound for us as well. Moreover, upon examining their statement and proof, we find that the value of $p$ used in \cite{DeligkasEGK024} was $k$ and hence the lemma simply gave an upper bound of $\bigoh(k^{3})$ on the number of moves, but we will set $p$ to be a value that is quadratic in $k$, hence we have restated the lemma to account for this. 
\end{remark*}

\begin{longlemma}\label{lem:implicationHaven}
Let $H_w$ be a strong $q$-haven anchored at some vertex $w$, where $q=\Theta(k^{2})$. 
Then, there exist three connected subgraphs $C_1, C_2, C_3$ of $G$ such that: (i) the pairwise intersection of the vertex sets of any pair of these subgraphs is exactly $w$, and (ii) $|V(C_1)| \geq k+1$, $|V(C_2)| \geq k+1$, and $|V(C_3)| \geq 2$, and (iii) $\bigcup_{i\in [3]} V(C_i) =V(H_w)$ and (iv) every vertex has distance $\bigoh(q)$ from $w$ in the graph induced by $\bigcup_{i\in [3]} E(C_i)$.
\end{longlemma}

\begin{proof}
   Say that $H_w$ is an $x$-$y$ path. Since $H_w$ is a strong $q$-haven anchored at $w$, it must be the case that     $w$ has distance $\bigoh(q)$ to both $x$ and $y$. 
   Let $x'$ and $y'$ denote the neighbors of $w$ on $H_w$, with $x'$ chosen as the one that is closer to $x$. 
   
   Since $w$ has degree at least three, there is a neighbor $u$ of $w$ that is distinct from $x'$ and $y'$. If $u$ is not in $H_w$, then we are done by choosing $C_1$ and $C_2$ to be the two subpaths from $w$ to the two endpoints of $H_w$ and choosing $C_3$ to be the graph comprising the edge $(u,w)$. 
   Otherwise, $u$ lies on $H_w$. Assume without loss of generality that $u$ lies between $w$ and $y$. Then, $u$ also lies between $y'$ and $y$. Then, set $C_1$ to be the subpath  $H_w[w,x]$. Set $C_3$ to be the graph comprising the edge $(w,y')$ and set $C_2$ to be the graph $H_w[w,y]+(w,u)-y'$.  

   This completes the proof of the lemma.
\end{proof}

For a strong $q$-haven $H_w$ anchored at some vertex $w$, we denote by $\hat H_w$, the graph obtained by taking the union of $H_w$ and an arbitrarily chosen set of three edges incident on $w$. 

\begin{longlemma}
\label{lem:swap}
Let $H_w$ be a strong $q$-haven anchored at some vertex $w$, where $q=\Theta(k^{2})$. 
For any set of robots $S \subseteq \R$ in $H_w$ with current configuration $\iota(S)$, any configuration $\iota'(S)$ with respect to $H_w$ can be obtained from $\iota(S)$ via a sequence of $\Oh(k^6)$ moves that take place in $\hat H_w$.
\end{longlemma}

\begin{proof}
We invoke Lemma \ref{lem:implicationHaven} followed by Lemma \ref{lem:oldLemSwap} with $p=q$. Notice that $\hat H_w$ contains the required neighbor $u$ of $w$ in the proof of Lemma \ref{lem:oldLemSwap}. Hence, the moves take place in $\hat H_w$. 
\end{proof}

A connected subgraph $G'$ is called an  $(r,q)$-{\em meta-haven} if it is obtained by taking the union of the edges of $\{\hat H_{w_i}\mid i\in [r]\}$, where each $H_{w_i}$ is a strong $q$-haven anchored at $w_i$. A straightforward consequence of Lemma \ref{lem:swap} is the following.

\begin{longlemma}
    \label{lem:metaSwap}
 Let $G'$ be an $(r,q)$-meta-haven, where $r\leq 2k$ and $q=\Theta(k^{2})$. 
For any set of robots $S \subseteq \R$ in $G'$ with current configuration $\iota(S)$, any configuration $\iota'(S)$ with respect to $G'$ can be obtained from $\iota(S)$ via a sequence of $\Oh(k^7)$ moves that take place in $G'$.
\end{longlemma}

We are now ready to complete the proof of the main lemma of this subsection. Let us restate it here. 

\stepbound*

\begin{proof}
If $G$ has treedepth bounded by $\bigoh(k^{2})$, then invoking  Lemma \ref{lem:mainTreedepth} with this bound on the treedepth, we conclude that if $I$ is a \YES-instance, then it has a schedule with  $2^{\bigoh(k^{4})}$ moves. 

Otherwise, Lemma \ref{lem:nearHaven} guarantees that for every vertex $v$, there is a $q$-haven for $v$ anchored at some vertex $w$,  where $q=\Theta(k^{2})$. Moreover, $H_w$ is a strong $q$-haven and has $v$ as one of its endpoints.
The proof idea from here onwards is similar to that of Eiben et al. (Theorem 15, \cite{DeligkasEGK024}). However, we have defined our notion of havens in such a way that our arguments are simpler. Assume without loss of generality that $R_1$ is a robot with a destination. Let us now describe a sequence of $\bigoh(k^{8})$ moves that take $R_1$ from $s_1$ to $t_1$. The complete schedule then repeats the same strategy for subsequent robots, giving us a bound of $\bigoh(k^{9})$ moves in total.

For every terminal $v$, fix $H_{w_v}$ to be a strong $q$-haven that has $v$ as one of its endpoints. 
Let $Z_1,\dots,Z_p$ denote meta-havens defined by $\hat H_{w_v}$, where $v$ is a terminal. By ensuring that each $Z_i$ is inclusion-wise maximal, we guarantee that $Z_i$ and $Z_j$ are disjoint for $i\neq j$. 

Let $P$ denote a walk from $s_1$ to $t_1$ in $G$. In the traversal of $P$ from $s_1$ to $t_1$, let $u_1$ and $u_2$ denote the first and last vertices, respectively, that belong to some meta-haven, say $Z_1$. Then, we replace the subwalk from $u_1$ to $u_2$ with a walk given by Lemma \ref{lem:metaSwap}, which takes $\bigoh(k^{7})$ moves. The other robots currently in the meta-heaven can be moved arbitrarily to other positions inside it. So, the only requirement of the configuration with which we invoke Lemma \ref{lem:metaSwap} is that $R_1$ starts in $u_1$ and ends in $u_2$. By doing this once for every meta-haven, we obtain a route for $R_1$ that enters and exits each meta-haven at most once and takes $\bigoh(k^{7})$ moves inside each meta-haven. Hence, we obtain a route for $R_1$ that takes $\bigoh(k^{8})$ moves overall. 
\end{proof}

\fi
\subsection{Succinct Representation of a Solution}

In what follows, fix an optimal schedule $\SSS=\{W_i\mid i\in [k]\}$ for an instance $I=(G, \R=(\M,\F), k, \ell)$. By Lemma \ref{lem:stepbound}, 
we may assume, without loss of generality, that $\SSS$ has length $2^{\bigoh(k^{4})}$.  
We also assume that no robot moves in two consecutive time steps since otherwise, the movements of such a robot  could be done in a single time step while the remaining robots wait. 
Moreover, we say that an edge $e$ of $G$ is {\em traversed by} $\SSS$ if there is an $i\in [k]$ and a time step $j$ such that the path $P^{i}_j$ contains the edge $e$. That is, some robot $R_i$ uses the edge $e$ in some time step $j$.  We denote by $G_\SSS$, the subgraph of $G$ induced by the edges traversed by $\SSS$. For a time step $j$, we denote by $G_{S,j}$ the subgraph of $G_\SSS$ induced by the edges traversed by $\SSS$ up to time step $j$.
A vertex $v$ in $G_\SSS$ is a \emph{waiting vertex} within time interval $[0,j]$ if there is a robot $R_i$ and a time step $j'\leq j$
% \todo{I: I think we should avoid using $t$ throughout this section (in the long version too). RMS: I've changed all occurrences.} 
such that $R_i$ is at $v$ at time steps $j'-1$ and $j$. We say that $v$ is a waiting vertex if it is a waiting vertex within some time interval. 
A vertex $v$ in $G_\SSS$ is an \emph{intersection} vertex within time interval  $[0,j]$ if its degree in $G_{S,j}$ is at least three. We say that $v$ is an intersection vertex if it is an intersection vertex within some time interval. 
A vertex $v$ in $G_{S,j}$ is an \emph{important} vertex within time interval $[0,j]$ if it is either a terminal vertex, or within time interval $[0,j]$ it is a waiting vertex or an intersection vertex. Otherwise, $v$ is \emph{unimportant} within time interval $[0,j]$.
We say that $v$ is an important (unimportant) vertex if it is important (resp.\ unimportant) within some time interval.

\begin{observation}
 
\label{boundWaitingbyTime}
 The number of waiting vertices within time interval $[0,j]$ is at most $k+j$.
\end{observation}

\iflong \begin{proof}
Initially, the robots are on $k$ terminals and in every time step, exactly one robot moves to a (potentially new) waiting vertex while the remaining robots remain stationary.
\end{proof} \fi

\begin{observation}
 
\label{disjointPaths}
The unimportant vertices within any time interval $[0,j]$ induce a disjoint union of paths and any endpoint of any of these paths has degree 2 in $G_{S,j}$.
\end{observation}

\iflong \begin{proof}
Every vertex of degree at least three in $G_{S,j}$ is important within this time interval.
\end{proof} \fi

\newcommand{\Pcal}{\mathcal P}

For a time step $j$, we denote by $\Pcal_{S,j}$, the set of all $x$-$y$ paths $P$ of non-zero length such that $x$ and $y$ are distinct important vertices within time interval $[0,j]$ and every internal vertex of $P$ is an unimportant vertex within time interval $[0,j]$.

For two paths $P$ and $P'$ in $G$, their {\em crossing points} are those vertices that have degree at least three in the graph induced by the edges in $E(P)\cup E(P')$. 

We say that $\SSS$ is a {\em special schedule up to time step $j$} if, for every robot $R_i$ and $j>0$ such that $R_i$ moves in time step $j$, the path $P^i_j$ has at most four crossing points with any path in $\Pcal_{S,j'}$ where $j'\leq j-1$. 
% Essentially, the path that $R_i$ takes when it moves in time step $j$ has at most four crossing points with any path that has been taken by any robot in any time step prior to time step $j$ and which is never blocked by a waiting vertex at any point up to time step $j-1$.\todo{I: This last sentence is long and confusing. Can it be rephrased/broken? RMS: I removed this. The formal description is already clear, I think.}
Note that every schedule is trivially special at time step 1 since only a single path has been taken at this point.  Moreover, $\Pcal_{S,1}$ has size at least one because a move has been made and so, $\Pcal_{S,j}$ has size at least one for every $j$. 
We say that $\SSS$ is a {\em special schedule} if it is a special schedule up to the final time step. 

We say that two schedules are {\em equivalent at time step} $j$ if the positions of all the robots at the end of time $j$ is the same in both schedules, that is, robot $R_i$ is on vertex $v$ after time step $j$ in one schedule if and only if the same happens in the other schedule.
We say that two schedules are equivalent if they are equivalent at every time step. In particular, two equivalent schedules take the same number of time steps.

%Iyad asked about whether we can provide an explanation for the proof of these, but for now we don't seem to have space and nobody felt like pushing for it.

\begin{lemma}
 
\label{lem:isSpecial}
	There is a solution $\SSS'$ that is a special schedule.
\end{lemma}

\iflong \begin{proof}
We start with the schedule $\SSS$ and suppose that it has $h$ time steps. Let $\SSS'$ be a schedule that is equivalent to $\SSS$ and let $j'$ be the maximum time step such that $\SSS'$ is a special schedule up to time step $j'$. We choose $\SSS'$ so that $j'$ is maximized. As observed above, $j'\geq 1$. Since we are done otherwise, we assume that $j'\neq h$. Now, let $j=j'+1$. 

 Suppose that $R_i$ moves in time step $j$ using the path $P^i_j$. By our choice of $j'$ (and by extension, $j$) it must be the case that some path
 $P$ in $\Pcal_{S',j-1}$
 and $P^i_j$ have more than four crossing points. Along the traversal of $P^i_j$ from $u^i_{j-1}$ to $u^i_j$, let the second and third crossing points be $x_1$ and $x_2$, respectively. Then, we can replace the subpath $P^i_j[x_1,x_2]$ with the subpath $P[x_1,x_2]$ to obtain an alternate schedule that is equivalent to $\SSS$ and with fewer crossing points.
 In fact, by doing the same for every path $P$ in $\Pcal_{S',j-1}$ that has more than two crossing points with $P^i_j$, we obtain a schedule $\SSS''$ that is equivalent to $\SSS$ and which is also special up to time step $j=j'+1$, a contradiction to our choice of $\SSS'$. This completes the proof of the lemma.
\end{proof} \fi

Henceforth, we assume that the solution $\SSS$ is a special schedule.

\begin{lemma}
 
\label{lem:boundImportant}
    The number of important vertices in $G_{\SSS}$ is bounded by $2^{\macrofk}$. 
\end{lemma}

\iflong \begin{proof}
Let $W$ denote the set of waiting vertices in $G_\SSS$ and let $I$ denote the set of intersection vertices in $G_\SSS$ that are not terminals or waiting vertices. Note that all but at most two of the terminals are by definition, waiting vertices. So, the number of important vertices is bounded by $|I|+|W|+2$.

By Observation \ref{boundWaitingbyTime},  since the total number of moves in $\SSS$ is at most $\macrofk$ (which also bounds the number of time steps in $\SSS$), the number of waiting vertices in $G_\SSS$ is at most $k+\macrofk$. 
 Therefore, it suffices to bound the size of $I$. Let $I[j]$ denote the subset of $I$ that is important within time interval $[0,j]$. 
 Clearly, $|I|\leq |\bigcup_{j\leq t }I[j]\setminus I[j-1]|$.
    Notice that every vertex in $I[j]\setminus I[j-1]$ is either a crossing point of the path $P^i_j$ with some path in $\Pcal_{S,j-1}$ or the endpoint of $P^i_j$ that is distinct from $u^{i}_{j-1}$.
    By Lemma \ref{lem:isSpecial}, we have that $P^i_j$ has at most four crossing points with any path in $\Pcal_{S,j-1}$. Hence, we have that $|I[j]\setminus I[j-1]|$ is at most $4\cdot |\Pcal_{S,j}|+1$. Moreover, since each path in $\Pcal_{S,j-1}$ gets split into at most five parts by the crossing points, we have that $|\Pcal_{S,j}|\leq 5\cdot |\Pcal_{S,j-1}|+1$. For simplicity, let us bound $5\cdot |\Pcal_{S,j-1}|+1$ by $6 \cdot |\Pcal_{S,j-1}|$ since $\Pcal_{S,j-1}$ is always non-empty. This implies that $|\Pcal_{S,j}|\leq 6^j$ and hence, $|I[j]\setminus I[j-1]|$ is at most $4\cdot (6^{j}+1)$. This gives us the required bound of $\bigoh(\macrofk\cdot 6^{\macrofk})=2^{\macrofk}$ on the size of $I$ and since $|W|\leq k+\macrofk$, completes the proof of the lemma.
\end{proof} \fi

\subsection{Leveraging Succinct Representations and Topological Minor Containment}

In what follows, fix a schedule $\SSS=\{W_i\mid i\in [k]\}$ for an instance $I=(G, \R=(\M,\F), k, \ell)$. 

\begin{figure}[ht]
    \begin{center}\hfill        \includegraphics[page=1]{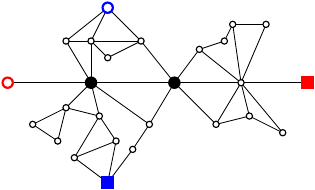} \hfill \includegraphics[page=2]{example.pdf}   \hfill~
    \end{center}
\caption{Left: an instance of \cmpl\ with two robots with destinations (red and blue, with destinations marked as squares) and two blocker robots (both marked as solid black circles). Right: A representation of one particular optimal schedule for this instance, where we first move the two blockers and then the remaining two robots.}
\label{fig:representation}
\end{figure}

We define the \emph{representation} of $\SSS$ to be the rooted graph (denoted by $H_\SSS$) whose vertices are the important vertices of $G_\SSS$, and in which there is an edge between two vertices $u$ and $v$ if and only if there is a $u$-$v$ path in $G_\SSS$ whose internal vertices have degree 2 in $G_\SSS$ and are unimportant. Moreover, the terminals receive unique labels (assume that the terminals are labeled with $[p]$ for some $p\leq 2k$) and the remaining vertices receive the same label which is distinct from those of the terminals, say $p+1$. An illustration is provided in Figure~\ref{fig:representation}.
From Lemma~\ref{lem:boundImportant}, we directly obtain: 

\begin{lemma}
\label{lem:representationbound}
If a solution exists, then there is a solution whose representation has $2^{\macrofk}$ vertices.
\end{lemma}

For the input graph $G$, we let $G'$ denote the rooted graph where the terminals get their unique labels and the rest of the vertices get a specific label $p+1$, where $p$ is the number of terminals.

\begin{lemma}
\label{lem:equivalence}
 $H_\SSS$ is a topological minor of $G$ and moreover, any realization of $H_\SSS$ as a topological minor in $G$ implies a schedule with length at most that of $\SSS$.
\end{lemma}

\begin{proof}
The forward direction is obvious and follows from the construction of $H_\SSS$.  

Consider the converse direction.
Notice that it is straightforward to  interpret $\SSS$ as a schedule on the graph $H_\SSS$ since the edges of $H_\SSS$ correspond to paths of $G_\SSS$ with no waiting vertices.
% we have only dissolved degree-2 vertices that are never waiting vertices. 
% 
Let $(\phi, \varphi)$ be a realization of $H_\SSS$ as a topological minor in $G$. We now define a schedule $\SSS'=\{W'_i\mid i\in [k]\}$ where each  
$W_i'=w_0^{i},Q_1^{i},\dots,Q_h^{i},w_h^{i}$. 
For every $i\in [k]$ and time step $j$, 
$w^{i}_j=\phi(u^{i}_j)$. Moreover, for $i\in [k]$ and time step $j$, if $P^{i}_j$ is a path with at least one edge then, $Q^{i}_j$ is obtained by concatenating the paths $\{\varphi(e)\mid e\in P^{i}_j\}$
in the natural way. The resulting schedule is equivalent to $\SSS$ by construction and comprises mutually non-conflicting routes since $\SSS$ is non-conflicting.  
\end{proof}

\begin{theorem}\label{thm:mainParbyBots}
    \cmpl{} is fixed-parameter tractable parameterized by the number of robots.
\end{theorem}

\begin{proof}
For a hypothetical optimal solution $\SSS$, we guess (i.e., use brute-force branching to determine) the representation $H_\SSS$. By Lemma \ref{lem:representationbound}, there are at most $r(k)$ choices for some $r(k)=2^{2^{\macrofk}}$ and these can be enumerated in FPT time. 
 For each guess $H_\SSS$, we do the following:
\begin{enumerate}
\item Decide whether the guess is feasible by brute-forcing over it. More precisely, we guess the configurations of the robots at each time step and verify that this guess corresponds to a set of non-conflicting routes with at most $\ell$ moves in total. \iflong Note that we can assume that no configuration repeats.\fi  Reject guesses that do not pass this check. Clearly, this step takes time $f(k)$ for some computable function $f$. 
\item For each guess $H_\SSS$ that is not rejected, invoke Proposition \ref{prop:TMcontainment} to decide whether $H_\SSS$ is a topological minor of $G'$ and return ``yes'' if and only if at least one of the invocations returns ``yes''.
\end{enumerate}

Fixed-parameter tractability follows from the fact that Proposition \ref{prop:TMcontainment} is used at most $f(k)$ times.
% \todo{I: The last lemma doesn't mention any running time. RMS: rephrased.} 
The dependence on $k$ in our algorithm for \cmpl{} is not made explicit since it relies on Proposition \ref{prop:TMcontainment} where this is also not explicitly stated. The correctness follows from Lemma \ref{lem:representationbound} and Lemma \ref{lem:equivalence}.
\end{proof}

\section{An \FPT{} Algorithm for Planar-CSMP-1 Parameterized by the Makespan}
\label{sec:planar}
We start this section by showing that \planaroneagentproblem{} is \NP-complete. The reduction is an adaptation of a reduction in~\cite{calinescu2008} for showing the \NP-hardness of a chip reconfiguration problem on grids:
\iflong
\begin{theorem}

\label{thm:nphardness}
	\planaroneagentproblem{} is \NP-complete.
 
\end{theorem}

\begin{figure}[ht]
    \begin{center} \includegraphics[page=1]{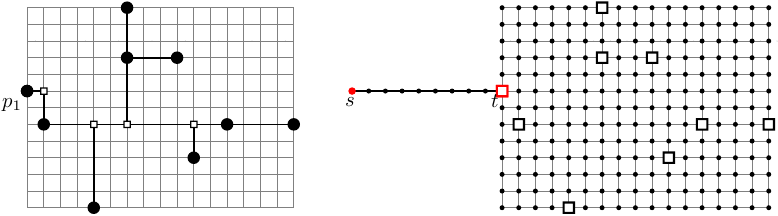} 
    \end{center}
\caption{An illustration for the reduction in Theorem~\ref{thm:nphardness}. The left figure shows an instance of STR, i.e., a set $P$ of points, and a Steiner tree for $P$. The right figure shows the corresponding instance of \planaroneagentproblem{}.}
\label{fig:nphard}
\end{figure}

\iflong
\begin{proof}
It is straightforward to see that the problem is in \NP. To show its \NP-hardness, we slightly modify the reduction in~\cite{calinescu2008}, which was used to prove the \NP-hardness of a related chip/coin moving problems on grids. In fact, we will show that the slice of the problem that consists of the restriction to instances in which the underlying graph is a subgrid (i.e., a subgraph of a rectangular grid), is \NP-hard. We closely follow the notations used in~\cite{calinescu2008}, and refer to Figure~\ref{fig:nphard} for illustration.

The reduction is from the Strongly \NP-hard problem Rectilinear Steiner Tree (RST)~\cite{GareyJohnson}. An instance of RST consists of a set $P=\{p_1, \ldots, p_n\}$ of $n$ points called \emph{terminals}. Without loss of generality, we will assume that $p_1$ is a leftmost point in $P$ and that its $x$-coordinate is zero. Since RST is Strongly \NP-hard, we can assume that all points in $P$ have integer coordinates and are encoded in unary. We are also given an integer $\ell \geq 0$, and the problem is to decide whether there is a Steiner tree, each of whose edges is either a horizontal or a vertical line segment, that connects all points in $P$ and such that the sum of the (Euclidean) lengths of its edges is at most $\ell$. 

To produce an instance of \planaroneagentproblem{}, we choose the smallest rectangular bounding box $B$ that encloses the points in $P$, and we construct a grid graph $G_B$ whose vertices are the points having integer coordinates in $B$ (including its boundaries). Observe that, we may assume that if the instance of RST has a solution, then it has a solution that is completely contained in $B$. Now we add to $G_B$ a path $(s, q_1, \ldots, q_{n-1})$ of new vertices that all lie on the same horizontal line as $p_1$, and such that the $x$-coordinate of $s$ is $-n$, and the $x$-coordinate of $q_i$ is $-n+i$, for $i=1, \ldots, n-1$.  We add the edge $(q_{n-1}, p_1)$, thus connecting the path to $G_B$ via $p_1$. This completes the construction of the plane graph $G$, which is a subgrid, in the instance of \planaroneagentproblem{}. The main robot in the instance is located at $s$ and its destination is $t=p_1$. All vertices of $G$ other than $s$ and the terminals are occupied by robots.  Clearly, the instance of \planaroneagentproblem{} can be constructed in polynomial time.

We prove the following claim (similar to the claim in Theorem 3.1 in~\cite{calinescu2008}), which establishes the correctness of the reduction: The instance of RST has a rectilinear Steiner tree of length 
at most $\ell$ if and only if the instance of \planaroneagentproblem{} has a solution that uses at most $\ell+1$ moves. 

To prove the direct implication, let $T$ be a solution to the instance of RST. Form the rooted tree $T'$ by adding to $T$ the path $(s, q_1, \ldots, q_{n-1}, p_1)$, and rooting $T'$ at $s$. We now apply the following process until eventually $s$ is located at $t=p_1$. We choose any leaf $x$ of $T'$ that is unoccupied at this point; we choose a closest (w.r.t.~distance in $T'$) occupied ancestor to $x$, move it to $x$ along $T'$, and remove $x$ from $T'$. Note that all the initially unoccupied vertices are in $T$ and after a vertex $x$ is processed, $x$ is occupied (before removal). Therefore, when we process $t$, all the vertices of $G_B$ are occupied and hence the main robot reached vertex $t$. We process every vertex of $T$ exactly once by moving a single robot once to that vertex. Hence, the total number of moves is equal to $|V(T)| = \ell +1$.  

To prove the converse, first observe that when the main robot reaches $t$, all other robots have to be in $G_B$. Consider all the paths induced by the moves of the robots in the solution. Observe that the union of these paths restricted to $G_B$ form a connected subgraph $H$ that includes all the terminals. Hence, $H$ contains a Steiner tree of $P$ of length at most $|V(H)|-1$.

Now, every vertex in $H$ is occupied at the end, and hence, it has to be an endpoint of a path-move. It follows that the number of moves is at least $|V(H)|$; that is $|V(H)|\le \ell+1$ and $H$ contains a Steiner tree of $P$ of length at most $\ell$.
\end{proof}
\fi

We now proceed to giving an \FPT{} algorithm for \planaroneagentproblem{} parameterized by $\ell$. We start by giving a high-level description of the algorithm.

Our aim is to prove that \planaroneagentproblem{} is fixed-parameter tractable parameterized by $\ell$. To do so, we will exploit the planarity of the graph to devise a reduction rule that, when repeatedly applied to an instance of the problem, either results in a simple instance that can be decided in polynomial time, or in an instance of bounded diameter, and hence of bounded treewidth (due to planarity). At the heart of this reduction rule is the identification of an edge which can be contracted without affecting the existence of a schedule. The crux of our technique is showing that a sufficiently large component of free vertices must contain an edge that can be safely contracted. These contractions eventually result in bounding the size of each free component in the graph, and consequently the treewidth of the graph. The fixed-parameter tractability of the problem will then be established using Courcelle's Theorem~(\Cref{fact:MSO}).

%At the heart of this reduction rule is a technique, the ``irrelevant edge'' technique~\cite{}\todo{I: Give a reference to papers that dubbed it as such. Do you know some prominent papers in which this terminology is used?}, which identifies an edge in the graph that can be safely contracted resulting in an equivalent instance of the problem. The crux of our technique is showing that a sufficiently large component of free vertices must contain an edge that can be safely contracted. These contractions eventually result in bounding the size of each free component in the graph, and consequently the treewidth of the graph. The fixed-parameter tractability of the problem will then follow using Courcelle's theorem (\Cref{fact:MSO}).
 
To show that each large component $C$ of free vertices contains an edge that can be safely contracted, we first show that this is the case unless $C$ is formed by a skeleton consisting of a sufficiently long path of degree-2 vertices (the degree is taken in $C$), plus a sufficiently small number of additional vertices. We then focus on the skeleton of this component. We show that if the instance is a \YES-instance of the problem, then it admits a solution $\SSS$ that interacts ``nicely'' with this skeleton. Before we give more details, recall from the previous section that a solution can be succinctly represented as a rooted topological minor of size $\sigma(\ell)$ for some computable function $\sigma$. This is because at most $\ell$ robots move in the solution.
Now, we can identify a subgraph $\SSS$ of the representation of the solution (i.e., a topological minor that is essentially a roadmap of the solution), that is separated from the rest of the solution by a small cut. The realization of this subgraph uses most of the skeleton and the remaining part of this realization consists of small subgraphs that are close to the skeleton. Therefore, we can assume that all edges in these parts of the subgraphs are realized by edges in the graph. 

Thus, it suffices to enumerate each possible representation of the part of the solution that interacts with the skeleton (which can be done in \FPT-time), and for each possible representation, test whether it exists near the skeleton; if it does, we mark the vertices in the graph that realize this representation. Since the skeleton is long, the total number of marked vertices on the skeleton is small, and an edge with both endpoints unmarked that can be contracted must exist. We remark that in many of the arguments that we make, we rely on the crucial fact that, except for the main robot, the robots are ``indistinguishable'' from each other.  We now proceed to the details. 

\smallskip
Refer to a vertex on which a robot other than the main robot is located as a \emph{blocker} vertex. 
Let the \emph{blocker-distance} between a vertex $v$ and a vertex $w$, denoted $\blockd(v,w)$, be the minimum integer $p$ such that there is a $v$-$w$ path in $G$ containing $p$ blocker vertices.
The reason why the blocker-distance is relevant is that we show that vertices which have high blocker-distance from $s$ and $t$ can be removed from the graph. 
Observe that, for any two vertices $v, w$,  $\blockd(v,w)$ can be computed in polynomial time via, e.g., Dijkstra's algorithm.
%\fi
%\iflong
%Observe that, for any two vertices $v, w$,  $\blockd(v,w)$ can be computed in $\Oh(n \lg{n})$ time using a shortest-path algorithm (e.g., Dijkstra's algorithm) on a planar graph of $n$ vertices. 
%\fi

% \todo{I: Is this lemma used?}
% \begin{lemma}
% If $\blockd(s,t)>\ell$, then $\III=(G,s,t,\ell,\R)$ is a \NO-instance.\todo{I: We're using a different notation for the instances than that in the problem definition and previous section.}
% \end{lemma}

% \iflong \begin{proof}
% A solution to the instance includes an $s$-$t$ walk (i.e., a sequence of paths leading the main robot from $s$ to $t$) that, by our assumption, must contain more than $\ell$ blockers in $G$ at time step 0. Since it requires at least one move to free up a walk from any particular blocker, freeing up any $s$-$t$ walk in $G$ from all the blockers it contains would require more than $\ell$ moves. 
% \end{proof} \fi

\iflong
The following observation is straightforward:

\begin{longobservation}
\label{obs:connected}
Let $\III=(G,s,t,\ell,\R)$ be a \YES-instance of the problem and consider a solution $\SSS$ of $\III$ of minimum makespan. Let $G_{\SSS}$ be the subgraph of $G$ consisting of all the paths in $\SSS$. Then $G_{\SSS}$ is connected and contains at most $\ell$ blockers in $G$.
\end{longobservation}

\begin{proof}
The latter claim follows immediately from the existence of a schedule with makespan $\ell$, while the connectivity of $G_{\SSS}$ follows from the fact that if it is disconnected, then it may contain at most a single connected component involving the moves of the main robot and we may obtain a new solution by removing all the other connected components (hence contradicting the minimality of $\SSS$).
\end{proof} \fi

%\begin{lemma}
%\label{lem:simpleprune}
%Let $v\notin \{s,t\}$ be vertex %such that $\max\{\blockd(v,s), %\blockd(v,t)\}>\ell$. If $v$ %contains a robot in $R$, let $R_v$ %be this robot. Then %$\III=%G,s,t,\ell,\R)$ is a \YES-%instance if and only if $(G-%v,s,t,\ell,\R \setminus \{R_v\})$ %is.
%\end{lemma}
 
% \iflong \begin{proof}
% Suppose that $\III$ is a \YES-instance and consider a solution $\SSS$ to $\III$ of minimum makespan. By Observation~\ref{obs:connected}, $G_{\SSS}$ is connected and contains at most $\ell$ blocker-vertices. Observe that $v \notin V(H)$: If $v \in V(H)$ then $\blockd(v,s) \leq \ell$ and $\blockd(v,t) \leq \ell$, and hence $\max\{\blockd(v,s), \blockd(v,t)\} \leq \ell$, contradicting our assumption. The lemma follows. 
% \end{proof} \fi

Let $\III=(G,s,t,\ell,\R)$ be an instance and let $F \subseteq V(G)$ be the set of free vertices in $G$.
A \emph{free component} is a connected component of $G[F]$, that is, a maximal connected subgraph of (initially) free vertices with pairwise blocker-distance $0$. Free components will be central to our proof---in particular, if the instance contains no ``large'' free components, then we can establish fixed-parameter tractability. To show this result, we first observe that the blocker-distance between $s$ and $t$ is at most $\ell - 1$, and hence, if there are no large free components, then the distance between $s$ and $t$ in $G$ is bounded. As the graph is planar, this also bounds the treewidth of the graph and we can use Courcelle's Theorem~\cite{Courcelle90,ArnborgLS91} to prove the following:

\begin{lemma}
 
\label{lem:smallfree}
\planaroneagentproblem can be solved in time $f(\ell,\lambda)\cdot |\III|$, where $f$ is a computable function and $\lambda$ is the size of the largest free component in the input instance $\III$. 
\end{lemma}
 
\iflong \begin{proof}
For the purposes of this proof, it will be useful to consider the main robot as a blocker as well.
By Observation~\ref{obs:connected}, we can assume that a solution $\SSS$ to $\III$ induces a connected subgraph $G_{\SSS}$ containing at most $\ell$ blockers. Any vertex in $G_{\SSS}$ is reachable (in $G_{\SSS}$) by a path from $s$. Any path $P$ in $G_{\SSS}$ that starts at $s$ has length $|P| \leq \ell \cdot (\lambda+1)$, since the subpath of $P$ between any two blockers belongs to a free component of $G$ of size at most $\lambda$ by our assumption. It follows that $G_{\SSS}$ is contained in the induced subgraph $G^-$ of $G$ centered at $s$ and of radius $\ell \cdot (\lambda+1)$. Therefore, we can reduce the instance to an equivalent instance in which the underlying graph is $G^-$. Since the diameter of $G^-$ is at most $2\ell \cdot (\lambda+1)$ and $G^-$ is planar, it follows that $tw(G^-) \leq 6 \ell \cdot (\lambda+1)+1$~\cite{RobertsonS84,biedl}.  

At this point, we can establish the lemma by constructing a suitable formula $\Phi$ in Monadic Second Order Logic (MSO$_2$) and invoking Courcelle's Theorem~\iflong
(\Cref{fact:MSO}). 
\fi 
Assume w.l.o.g.\ that $\ell\geq 2$, as the case where $\ell=1$ can be solved in polynomial time. To construct our formula $\Phi$, we inductively construct the following $\ell$ subformulas $\texttt{Move}_{1},\dots, \texttt{Move}_{\ell}$:  

\begin{itemize}
\item $\texttt{Move}_{1}(M_1)$ has a single free vertex set variable $M_1$ and is true if and only if all the following conditions hold:
\begin{itemize}
\item[(a)] $M_1$ forms a path $P_1$ and moreover;
\item[(b)] $M_1$ contains precisely one vertex $s_1$ which is not free (i.e., $s_1$ is either $s$ or an occupied vertex); and
\item[(c)] $s_1$ is an endpoint of $P_1$.
\end{itemize}
\end{itemize}

Note that if $\texttt{Move}_{1}(M_1)$ is true, then it yields a uniquely defined choice of $s_1$; we denote the other endpoints of $M_1$ as $t_1$ and note that both vertices can be expressed using MSO$_2$ logic based on $M_1$. 
To proceed, we construct the subformula $\texttt{Free}_1(v,M_1)$ which is true if and only if the vertex $v$ is either $s_1$, or $v$ is free in $G^--t_1$ (intuitively, this formula captures the property of ``being free after moving a robot according to $M_1$''). Before providing the full iterative definition of these notions, we make the second $\texttt{Move}$ subformula explicit below:

\begin{itemize}
\item $\texttt{Move}_{2}(M_1,M_2)$ has two free vertex set variables $M_1$, $M_2$ and is true if and only if all the following conditions hold:
\begin{itemize}
\item[(0)] $\texttt{Move}_{1}(M_1)$ is true;
\item[(a)] $M_2$ forms a path $P_2$;
\item[(b)] $M_2$ contains precisely one vertex $s_2$ such that $\texttt{Free}_1(s_2,M_1)$ is not true; and
\item[(c)] $s_2$ is an endpoint of $P_2$.
\end{itemize}
\end{itemize}

Now, assume we have inductively defined subformula $\texttt{Move}_{i}(M_1,\dots,M_i)$, where we require $M_i$ to be a path with designated endpoints $s_i$ and $t_i$. Furthermore, assume we have also constructed a subformula $\texttt{Free}_{i-1}(v,M_1,\dots,M_{i-1})$. We define the subformula $\texttt{Free}_i(v,M_1,\dots,M_i)$ as follows: the subformula is true if and only if the vertex $v$ is either $s_i$, or $\texttt{Free}_{i-1}(v,M_1,\dots,M_{i-1})$ holds in $G^--t_i$. Then we set:

\begin{itemize}
\item $\texttt{Move}_{i+1}(M_1,\dots,M_i)$ has $i+1$ free vertex-set variables $M_1$, \dots, $M_{i+1}$ and is true if and only if all the following conditions hold:
\begin{itemize}
\item[(0)] $\texttt{Move}_{1}(M_1)$, \dots, $\texttt{Move}_{i}(M_1,\dots,M_{i})$ are all true;
\item[(a)] $M_{i+1}$ forms a path $P_{i+1}$;
\item[(b)] $M_{i+1}$ contains precisely one vertex $s_{i+1}$ such that $\texttt{Free}_{i}(s_{i+1},M_1,\dots,M_i)$ is not true; and
\item[(c)] $s_{i+1}$ is an endpoint of $P_{i+1}$.
\end{itemize}
\end{itemize}

To complete our description, we iteratively define a subformula to keep track of the location of the main robot after each move: 
\begin{itemize}
\item $\texttt{MainRobot}_0(v)$ is true if and only if $v=s$; and
\item for each $1\leq i\leq \ell$, $\texttt{MainRobot}_i(v)$ is true if and only if $(\texttt{MainRobot}_{i-1}(v)\wedge (v\neq s_i)) \vee (\texttt{MainRobot}_{i-1}(s_i)\wedge (v=t_i))$. In other words, the main robot is located at $v$ in step $i$ if and only if either $v$ was the location of the main robot after the previous move and the current move did not change that, or $v$ is the endpoint of the last move and the last move did move the main robot.
\end{itemize}

Finally, we set $\Phi(M_1,\dots,M_\ell)= \texttt{Move}_{\ell}(M_1,\dots,M_\ell) \wedge (t_\ell=t) \wedge \texttt{MainRobot}_\ell(s_\ell)$.

For correctness, assume that there exists a tuple $(X_1,\dots,X_\ell)$ of vertex subsets satisfying $\Phi(X_1,\dots,X_\ell)$. Then we can construct a schedule as follows. First, we move the robot with position $s_1$ to $t_1$. Then, we move the robot which at this point has position $s_2$ to $t_2$, and so forth, until we complete $\ell$ moves. For each such move, the formula guarantees that the path taken by the robot cannot be occupied, and moreover our conditions ensure that the last move brings the main robot to its destination. We note that the adopted assumption of having the final $\ell$-th move concern the main robot is without loss of generality: if the main robot is already at its destination after the previous move, the formula allows one to simply set $X_\ell=\{t\}$ to model ``skipping a move''.

For the converse direction, assume that $G^-$ admits a schedule consisting of at most $\ell$ moves. Since each $i$-th move is a path $P_i$ containing precisely one vertex that is not free at time step $i$, setting each $M_i$ to be precisely the vertices of $P_i$ will satisfy the formula. Since the construction of $\Phi(M_1,\dots,M_\ell)$ can clearly be done in time polynomial in $\ell$, the lemma indeed follows by applying \Cref{fact:MSO} on the constructed formula.
\end{proof} \fi

To apply Lemma~\ref{lem:smallfree}, we will devise a non-trivial reduction procedure that will eventually guarantee a bound on the size of each free component in $\III$.

\subparagraph{The Structure of Free Components.}
%Observe that for any three vertices %$x, c_1, c_2$ such that $c_1,c_2\in %C$, $\blockd(x,c_1)=\blockd(x,c_2)$. 
%Let $\blockd(s,C)$ ($\blockd(t,C)$) be the blocker-distance between $s$ ($t$, respectively) and an arbitrary vertex in $C$. 
The following observation is straightforward:

\begin{observation}
\label{obs:resilient}
    Let $Q$ be a path of free vertices and let $x,y$ be arbitrary vertices on $Q$ such that the subpath of $Q$ between $x$ and $y$ has length at least $\ell$. If there exists an $x$-$y$ path $P_{xy}$ whose internal vertices are disjoint from $Q$ such that the number of blockers on $P_{xy}$ is at most $\ell - \blockd(s,x) - \blockd(y,t) - 1$, then the instance $\III$ is a \YES-instance and we can output a solution for $\III$ in polynomial time. 
\end{observation}

\iflong
\begin{proof}
We can construct a solution by moving all blockers on $P_{xy}$, all blockers on an $s$-$x$ path witnessing $\blockd(s,x)$ and all blockers on an $y$-$t$ path witnessing $\blockd(y,t)$, into the $x$-$y$ subpath of $Q$. Afterwards, we can use a single move to transfer the main agent from $s$ to $t$ along an unobstructed path.
\end{proof}
\fi

We refer to a path where the above observation does not result in solving the instance as a \emph{resilient} path. In other words, a path $Q$ is resilient if and only if for every pair of vertices $x, y$ in $V(Q)$ whose \iflong $Q$-distance\fi is at least $\ell$, there is no $x$-$y$ path $Z$ in $G-Q$ with at most $\ell- \blockd(s,x) - \blockd(y,t) -1$ blockers.

Our goal is to show that every ``sufficiently large'' free component contains a ``still sufficiently long'' path with certain special properties that, intuitively, allow the path to be ``cleanly separated'' from a hypothetical solution. This is formalized in the following definition, where the property of being ``still sufficiently long'' is given by the function $g(\ell)= 32 \ell^6 \cdot 2^{14 \ell^2}+1$.  
\iflong
The reason for this choice of $g(\ell)$ will become clear later (in the proof of Lemma~\ref{lem:reductionsafe}).
\fi

\begin{definition}
    \label{def:clean}
Let $Q$ be a path in a free component $C$ with endpoints $u'$, $v'$. We say that $Q$ is $\ell$-\emph{clean} if it satisfies the following conditions:
\begin{enumerate}
\item every vertex in $Q$ has degree precisely $2$ in $C$;
\item $|V(Q)|> g(\ell)$;

\item if $\mathcal{I}$ admits a solution that intersects $V(Q)$, then there are two vertices $u,v\in V(Q)$, each at distance at most $\ell^2+1$ from $u'$ and $v'$, respectively, and a solution $\SSS$ for $\mathcal{I}$ such that:
 
 $\{u,v\}$ is a cut in $G_{\SSS}$ between $\{s,t\}$ and the $u$-$v$ subpath of $Q$; we denote by $Q^{\SSS}_{uv}$ the connected subgraph of $G_{\SSS}$ that contains $u$, $v$ and all the connected components of $G_{\SSS}-\{u,v\}$ that do not contain $s$ or $t$; and
    
\item $Q$ is resilient.
\end{enumerate}
\end{definition}

\iflong
\begin{longlemma}
\label{lem:structure}
    Let $\III$ be an instance of \planaroneagentproblem{}, and let $C$ be a free component of size at least $3\ell$. In polynomial time, we can output one of the following: 
    \begin{itemize}
        \item A solution $S$ that intersects $C$; 
        \item a decision that no solution for $\III$ intersects $C$; or
 
        \item a path $Q$ in $C$ of length at least $|C| - \ell + 2$ with endpoints $p,q$ such that 
        \begin{itemize}
            \item $Q$ is a shortest $p$-$q$ path in $C$, and
            \item for any solution $\SSS$, if for vertices $x, y$ on $Q$ the graph $G_{\SSS}-(V(Q)\setminus \{x,y\})$ contains an $s$-$x$ path and a $t$-$y$ path, then $\dist_{Q}(x, p) + \dist_{Q}(y, q) \leq \ell-2$.  
        \end{itemize}
        % for any solution $\SSS$, if for vertices $x, y$ on $Q$ the graph $G_{\SSS}-(V(Q)\setminus \{x,y\})$ contains an $s$-$x$ path and $t$-$y$ path, then the $Q$-distance of $x$ and $p$ plus the $Q$-distance of $y$ and $q$ is at most $\ell$.
           \end{itemize}
\end{longlemma}
\begin{proof}
    Suppose that there exists a solution $\SSS$ that intersects $C$. Let $p$ and $q$ be two vertices (not necessarily distinct) in $C$ such that $G_{\SSS} - (V(C)\setminus\{p\})$ contains an $s$-$p$ path $P_{sp}^{\SSS}$, and $G_{\SSS} - (V(C)\setminus\{q\})$ contains a $q$-$t$ path $P_{qt}^{\SSS}$. Let $Q_{pq}$ be a shortest path between $p$ and $q$ in $C$. 
    If $V(C) \setminus V(Q_{pq})$ contains at least $\ell-1$ vertices, then it is easy to see that there exists a solution $\SSS'$ that starts by moving all the blockers in $V(P_{sp}^{\SSS}) \cup V(P_{qt}^{\SSS})$ to the vertices in $V(C) \setminus V(Q_{pq})$, and then moves the main robot from $s$ to $t$ in a single move along the edges of $P_{sp}^{\SSS}\cup Q_{pq}\cup P_{qt}^{\SSS}$. We now show that, for every pair $p,q \in V(C)$, we can check the existence of a solution $\SSS'$ w.r.t.\ $(p,q)$ in polynomial time. We distinguish two cases depending on whether $P_{sp}^{\SSS}$ is vertex-disjoint from $P_{qt}^{\SSS}$ or not. 
    
    First, if $P_{sp}^{\SSS}$ and $P_{qt}^{\SSS}$ are disjoint, then $\blockd(s,p)+\blockd(q,t)\le \ell-1$. Hence, we first compute the distance between $p$ and $q$ in $C$ and if it is less than $|C|-\ell + 1$, then we compute a shortest path (w.r.t.~the blocker-distance) from $s$ to $p$ and a shortest path from $p$ to $t$, and construct the solution $\SSS'$ as discussed above.

    Second, if $P_{sp}^{\SSS}$ and $P_{qt}^{\SSS}$ intersect, then it is easy to verify that there exists a vertex $y\in V(G_{\SSS})- V(C)$ such that $P_{sp}^{\SSS} \cup P_{qt}^{\SSS}$ contains three paths $P_{sy}$ from $s$ to $y$,  $P_{yp}$ from $y$ to $p$, and $P_{yt}$ from $y$ to $t$ such that the three paths pairwise  intersect only in $y$. It follows that $\blockd(s,y) + \blockd(y,t) + \blockd(y,p)\le \ell -1$. Therefore, 
    similarly to the process above, in polynomial time, by enumerating/trying each vertex in the graph as $y$ and computing shortest paths in $G$ from $s$-$y$, $y$-$t$, and $y$-$p$, we can find three paths such that the total number of blockers on them is at most $\ell-1$. Note that $P_{yp}$ is vertex disjoint from $s$ and we actually need to find a $y$-$p$ path that does not contain $s$, but that is just finding a shortest path in $G-s$. We now move the blockers on these paths, using at most $\ell-1$ moves to $C$, followed by a move of the main robot from $s$ to $t$. 
    
    Notice that from the second case, it also follows that if $\SSS$ intersects $C$ and $G_{\SSS}$ contains an $s$-$t$ path that is disjoint from $C$, then the above enumeration would find this solution. Therefore, we can assume that $C$ is an $s$-$t$ cut in $G_{\SSS}$, and we can assume that $P_{sp}^{\SSS}$ and $P_{qt}^{\SSS}$ are vertex disjoint. Hence $\blockd(s,p) + \blockd(q,t) \le \ell -1$ and $p$ and $q$ are at distance at least $|C|-\ell +2$ in $C$. Now, there has to be at least one pair of vertices $p,q$ in $C$ such that $\blockd(s,p) + \blockd(q,t) \le \ell -1$, and that they are at distance at least $|C|-\ell +2$; otherwise, we either have produced a solution as in the above, or no solution intersects $C$ and we report that. Suppose now that a pair $p, q$ satisfying the above inequality exists. We argue that we can now output a shortest path, $Q$, between $p$ and $q$ in $C$. From the above, $Q$ has length at least $|C|-\ell+2$ and is a shortest $p$-$q$ path in $C$. It only remains to show that no solution $\SSS$ can contain a pair of vertices $x,y$ on $Q$ such that the graph $G_{\SSS}-(V(Q)\setminus \{x,y\})$ contains an $s$-$x$ path and a $t$-$y$ path satisfying $\dist_{Q}(x, p) + \dist_{Q}(y,q) \geq \ell-1$. 
     However, if this were the case then the shortest $x$-$y$ path would have length at most $|Q|-\ell +1 \le |C|-\ell+1$, and hence we would have produced a solution during the above enumerations.
\end{proof} \fi

\begin{lemma}
 
\label{lem:freepath}
There is a polynomial-time procedure which takes as input a free component $C$ in an instance $\III=(G,s,t,\ell,\R)$ such that $|V(C)|\geq (g(\ell)+1) \cdot (\ell-1)+3(\ell+2)$, and either solves the instance, or outputs an instance equivalent to $\III$ obtained by contracting an edge in $C$, or outputs an $\ell$-clean path $Q$ in $C$.
\end{lemma}

\iflong \begin{proof}
Consider a free component $C$ such that $|V(C)|\geq (g(\ell)+1) \cdot (\ell-1)+3(\ell+2)$. 
By Lemma~\ref{lem:structure}, we can either find a solution that intersects $C$, determine that a solution cannot intersect $C$, or find a shortest path $Q'$ in $C$ between two vertices $p$ and $q$ of length at least $|C|-\ell +1$. 
Moreover, for every solution $\SSS$ that intersects $C$, and for any two vertices $x$ and $y$ such that $G_{\SSS}$ contains an $s$-$x$ path $P_{sx}$ and a $y$-$t$ path $P_{yt}$ that are internally vertex-disjoint from $Q'$, it holds that $\dist_{Q'}(p,x)+\dist_{Q'}(q,y)\le \ell-2$. Since $Q'$ is a shortest $p$-$q$ path in $C$, every vertex $z\in V(C) \setminus V(Q')$ has at most two neighbors on $Q'$, and if it has two neighbors, these neighbors appear consecutively on $Q'$ (otherwise, $Q'$ could be shortcut). 
Since, $|V(C) \setminus V(Q')| \le \ell - 2$, $Q'$ can be split into at most $\ell-1$ disjoint subpaths, each consisting of vertices of degree $2$ in $C$. Hence, one of these subpaths, $Q$, must have length at least $\lceil\frac{|V(C)|-3(\ell+2)}{\ell-1}\rceil$. 
Let $u', v'$ be the endpoints of this path such that $u'$ is closer to $p$ and $v'$ is closer to $q$. Note that, for any $x\in V(Q)$, it holds that $\dist_{Q}(u',x) = \dist_{Q'}(u',x) \le \dist_{Q}(p,x)$ and $\dist_{Q}(v',x) = \dist_{Q'}(v',x)\le \dist_{Q}(q,x)$. Hence, for any solution $\SSS$
that intersects $Q$ and any two vertices $x$ and $y$ such that $G_{\SSS}$ contains an $s$-$x$ path $P_{sx}$ and a $y$-$t$ path $P_{yt}$ that are internally vertex-disjoint from $Q$, it holds that $\dist_{Q}(u',x)+\dist_{Q}(v',y)\le \ell-2$.
 
  The path $Q$ satisfies properties 1 and 2 of Definition~\ref{def:clean} of an $\ell$-clean path. Moreover, by Observation~\ref{obs:resilient}, we may assume that $Q$ is resilient, and this satisfies property 4 of Definition~\ref{def:clean}. In what follows, we will show that we can assume that $Q$ satisfies property 3 of Definition~\ref{def:clean}, or otherwise we can output a smaller equivalent instance to $\III$. 

To prove property 3, assume that $\III$ admits a solution $\SSS$ that intersects $Q$ and note that $G_{\SSS}$ is connected by Observation~\ref{obs:connected}. Let $u', v'$ be the endpoint of $Q$. 

%By Lemma~\ref{lem:structure} and selection of $Q$, we may assume that $\SSS$ satisfies the following property. For every vertex $x$ on $Q$ such that there is an $s$-$x$ path in $G_{\SSS} - Q$, the $Q$-distance between $u'$ and $x$ is at most $\ell$. 
By the selection of $Q$ and the discussion above, we can assume that, for every vertex $x$ on $Q$ such that there is an $s$-$x$ path in $G_{\SSS} - Q$, the $Q$-distance between $u'$ and $x$ is at most $\ell-2$ and similarly, for every vertex $y$ on $Q$ such that there is a $t$-$y$ path in $G_{\SSS} - Q$, the $Q$-distance between $v'$ and $y$ is at most $\ell-2$.

Now, let us consider the vertices $u_1, u_2, \ldots, u_\ell$ on $Q$ such that, for each $r\in [\ell]$, the vertex $u_r$ is the vertex with $Q$-distance $r\cdot \ell + 1$ from $u'$ on $Q$. 
Similarly, let us consider the vertices $v_1, v_2, \ldots, v_\ell$ on $Q$ such that, for each $r\in [\ell]$, the vertex $v_r$ is the vertex with $Q$-distance $r\cdot \ell + 1$ from $v'$ on $Q$. We claim that there is a pair $(u_i, v_j)$ that satisfies property 3. 
First, note that for every vertex $x\in X=\{u_1,\ldots, u_\ell, v_1,\ldots, v_\ell\}$, there is no path in $G_{\SSS} - (V(Q)\setminus \{x\})$ from $s$ nor $t$ to $x$, because $x$ is at $Q$-distance at least $\ell$ from both $u'$ and $v'$.
If $\{u_i, v_j\}$, for some $i,j\in [\ell]$, is not a cut in $G_{\SSS}$, then $G_{\SSS}$ contains a path between two vertices $a,b$ such that $a$ is on $Q$ between $u_i$ and $v_j$ and $b$ is on $Q$ outside of the $u_i$-$v_j$ subpath of $Q$; we refer to such an $a$-$b$ path as a \emph{detour}. 
Since $Q$ is resilient, the $Q$-distance between $a$ and $b$ in a detour is at most $\ell-1$, and hence the $a$-$b$ subpath of $Q$ contains at most one of the vertices in $X$.
Moreover, any two detours in $G_{\SSS}$ for which the subpaths contain different vertices in $X$ are vertex disjoint (otherwise they can be merged into a detour that connects two vertices on $Q$ with $Q$-distance at least $\ell$) and every such detour contains a blocker (since the vertices in $Q$ have degree 2 in $C$). 
Hence, there are at most $\ell-1$ such detours and there exists a $\{u_i, v_j\}$ pair such that no detour contains $u_i$ nor $v_j$. It follows that $\{u_i, v_j\}$ cuts $\{s,t\}$ in $G_{\SSS}$ from the subpath of $Q$ between $u_i$ and $v_j$. 
% \todo{I: This is leftover, no?}
% \textcolor{red}{Let $u$ be the vertex in $V(G_{\SSS}) \cap V(Q)$ whose $Q$-distance from $u'$ is maximized, and such that there is an $s$-$u$ path in $H$ $v$ the vertex in $V(H) \cap V(Q)$ that is farthest to $v'$. 
% Let $Q_{u'u}$ be the subpath of $Q$ between $u'$ and $u$. If $|Q_{u'u}| \geq \ell$, then since $\SSS$ is connected by Observation~\ref{obs:connected}, $\SSS$ can be modified into a solution that starts by moving the at most $\ell-1$ blockers on $\SSS$ into the vertices on $Q_{u'u}|$ excluding $u$, and then follow that by a single move of the main robot along $S$ from vertex $s$ to vertex $t$. Such a modified solution clearly won't be affected by contracting an edge of $Q$ since it would leave at least $\ell-1$ free vertices excluding $u$ on $Q_{u'u}$. Therefore, we can assume that $u$ is of distance at most $\ell$ from $u'$. Similarly, we can assume that $v$ is at distance at most $\ell$ from $v'$.} 
\end{proof} \fi

\subsection{Finding Irrelevant Edges}

For the following considerations, it will be useful to proceed with a fixed choice of $C$ and an $\ell$-clean path $Q$. Next, we will show that we can make an even stronger assumption about the hypothetical solutions ``passing through'' $Q$.

\begin{lemma}
 
\label{lem:pendingvertices}
Assume that $\mathcal{I}$ admits a solution that intersects $V(Q)$. Then there exist vertices $u$, $v$ each at distance at most $\ell^2+1$ from $u'$ and $v'$, respectively, and a solution $S^*$ for $\III$ satisfying properties 3-4 in Definition~\ref{def:clean}, and furthermore satisfying the following property 5: $|V(Q^{S^*}_{uv}) - V(Q)|\leq \ell$.
\end{lemma}

\iflong \begin{proof}
By Lemma~\ref{lem:freepath}, we can assume that $\III$ has a solution $\SSS$ that satisfies property 3 in the definition of $\ell$-clean path. Let $u', v'$ be the endpoints of $Q$, and $u, v, Q_{uv}^{\SSS}$ be as in property 3 of the definition. Assume further that $|V(Q^{\SSS}_{uv}) - V(Q)| > \ell$, and we will show how a solution $\SSS^*$ can be obtained that satisfies properties 1-5.

Since $|V(Q^{\SSS}_{uv}) - V(Q)| > \ell$ and $Q^{\SSS}_{uv}$ is connected to $Q$, we can find a set $Y$ of $\ell$ vertices in $V(Q^{\SSS}_{uv}) - V(Q)$ such that the subgraph $Q_{uv}^{*}$ consisting of the path $Q_{uv}$, the subgraph of $Q_{uv}^{\SSS}$ induced by $Y$, and the edges between $Q_{uv}$ and the subgraph of $Q^{\SSS}_{uv}$ induced by the vertices in $Y$ is connected. Now consider the subgraph $G_{\SSS^*}$ of $G$ obtained from $G_{\SSS}$ by replacing $Q^{\SSS}_{uv}$ with $Q_{uv}^{*}$. It is easy to see that $G_{\SSS^*}$ is connected and $G_{\SSS^*} - Y$ contains an $s$-$t$ path. 
We can define a solution $\SSS^*$ whose corresponding subgraph is $G_{\SSS^*}$ as follows. Observe that the number of blockers in $G_{\SSS^*}$ is at most $\ell -1$ and that $Y$ has $\ell$ vertices (some of them are possibly blockers). We start by moving all blocker vertices in $G_{\SSS^*}$ to $Y$ possibly pushing some of the blockers in $Y$ to other (free) vertices in $Y$ along the way; this costs at most $\ell-1$ moves. We then follow that by a move of the main robot along the free $s$-$t$ path in $G_{\SSS^*} - Y$. Clearly, all properties 3-5 are now satisfied.
\end{proof} \fi

For the following, we will refer to a solution satisfying properties 3-5 as $(Q,u,v)$-\emph{canonical}, or simply $Q$-canonical in brief when the identities of $u$ and $v$ do not matter. From the properties of canonical solutions, we can directly establish the following:
%\todo{I: Can we use a different font for ``waiting'' and ``occupied''} R: We changed the occupied label, but waiting seems to be just an adjective so didn't change it for now.

\begin{lemma}
 
\label{lem:template}
Let $S^*$ be a $Q$-canonical solution.
Then there is a $(2\ell^2)$-vertex rooted graph $U^{S^*}_{uv}$ (with roots $u$, $v$ and \texttt{occupied}), such that $U^{S^*}_{uv}$ is a rooted topological minor of $G$ (with roots $u$, $v$ and \texttt{occupied}) and there is a realization $(\phi, \psi)$ of $U^{S^*}_{uv}$ that maps $U^{S^*}_{uv}$ to $Q^{S^*}_{uv}$.
%\todo{$Q^{S^*}_{uv}$ is topological minor model of $U^{S^*}_{uv}$; TODO: define topological minor model as subgraph $G'$ of $G$ is topological minor model of $H$ if ... \\ RMS: Added to prelims}. 
Moreover, waiting vertices are in $\phi(U^{S^*}_{uv})$ and for every edge $e\in E(U^{S^*}_{uv})$ either $\psi(e)$ is a single edge or a subpath of $Q$. 
% \todo{We should ask RMS about terminology.}
%Then $Q^{S^*}_{uv}$ can be obtained by subdividing a $(2\ell)$-vertex graph $U^{S^*}_{uv}$, where every vertex of $U^{S^*}_{uv}$ may either be labeled as occupied or free, every vertex obtained via subdivisions is free, and there is a unique $u$-$v$ path which moreover contains all of the vertices obtained via subdivisions.
\end{lemma}

\iflong \begin{proof}
The lemma trivially follows from Lemma~\ref{lem:pendingvertices} and the resiliency of $Q$. There are at most $\ell-1$ vertices in $V(Q^{S^*}_{uv}\setminus Q)$ and each has at most $\ell-1$ neighbors in $Q$. All other vertices are on a subpath of $Q$ between two vertices on $Q$. Of these at most $\ell-1$ can be waiting vertices. 
\end{proof} \fi

Crucially, below we provide a procedure that can efficiently test for $U^{S^*}_{uv}$ as defined above.
%, but to do so we will need to establish a bound on the treewidth of a suitable subgraph of $G$. Towards this, we prove the following general lemma.

%We can now state the testing result we will later use to identify ``safe'' edges to contract on $Q$. \todo{I: This sentence seems out of place given the last sentence in the paragraph above it.}

\begin{lemma}
 
\label{lem:test}
Let $Q$ be an $\ell$-clean $u'$-$v'$ path and let $u$, $v$ be the vertices specified in property 3 of Definition~\ref{def:clean}. 
Let $p\leq 2\ell$ and $U$ be a connected $p$-vertex rooted graph with roots $u$, $v$ and ``\texttt{occupied''} such that $U$ contains at most $\ell - \blockd(s,x) - \blockd(y,t) - 1$ occupied vertices. There is a procedure which runs in time $2^{\bigoh(\ell^2)}\cdot |V(G)||V(Q)|$ and either outputs a realization $(\phi, \psi)$ of $U$ which results in a subgraph $\Gamma\subseteq G$ such that:
%outputs a subgraph $\Gamma\subseteq G$ and a realization $(\phi, \psi)$ of $U$ which results in $\Gamma$ such that:
\begin{itemize}
\item $u,v$ separate $\Gamma$ from $u'$ and $v'$, and 
\item for every edge $e\in E(U)$ either $\psi(e)$ is a single edge or a subpath of $Q$, or
\end{itemize}
correctly determines that no such $(\Gamma,(\phi,\psi))$ exists. 
%Moreover, if such $(\Gamma,(\phi,\psi))$ exists, then $|N(\Gamma-Q)\cap Q|\leq p\cdot\ell$.
\end{lemma}

\iflong \begin{proof}
%First, observe that the conditions on $\Gamma$ guarantee that it must lie completely at distance at most $\ell$ from $Q$. Moreover, 
Let us assume that there exist $(\Gamma,(\phi,\psi))$ satisfying the conditions of the lemma. First, note that for each connected component $C$ of $\Gamma-Q$, every pair of vertices in $N(C)\cap Q$ has $Q$-distance at most $\ell$. Indeed, by our assumptions we have that $C$ contains at most $\ell - \blockd(s,x) - \blockd(y,t) - 1$ occupied vertices, as by $\psi$ we are not allowed to contract edges outside of $Q$. Since $Q$ is $\ell$-clean (and in particular resilient), we know that there is no path $Z\in G-Q$ with at most $\ell- \blockd(s,x) - \blockd(y,t) -1$ blockers between any pair of vertices on $Q$ with $Q$-distance greater than $\ell$. Hence, every pair of vertices in $N(C)\cap Q$ must have $Q$-distance at most $\ell$, as desired. %This establishes the last sentence of the lemma.

For the next step, we recall that the conditions placed on $(\Gamma,(\phi,\psi))$ guarantee that each connected component $C$ of $\Gamma-Q$ has size at most $p\leq 2\ell$, and that each such $C$ is isomorphic to some induced subgraph $C'$ of $U$. Moreover, each $(\Gamma,(\phi,\psi))$ with connected components $C_1,\dots,C_i$ of $\Gamma-Q$ yields a unique partition $\Upsilon$ of $V(U)$ into (1) vertex sets $U_1,\dots,U_i$, each of which induces a subgraph isomorphic to the corresponding connected component in $C_1,\dots,C_i$, and (2) the set of vertices $U^*$ which are mapped by $\psi$ to $Q$. Armed with this knowledge, we perform exhaustive branching to consider all options for the following information:

\begin{enumerate}
\item a partitioning of $U$ into $U^*, U_1,\dots,U_i$ (for every possible choice of $i\leq p$);
\item an ordering $\lambda$ of the vertices in $U^*$; and
\item for each pair of vertices $a,b$ which are consecutive in $\lambda$, a value $\texttt{U-dist}(a,b)$ from $[\ell]\cup \{\texttt{large}\}$.
\end{enumerate}

As stated above, each $(\Gamma,(\phi,\psi))$ satisfying the conditions of the lemma yields a unique partitioning as per point $1$; moreover, it also uniquely determines an ordering $\lambda$ of $U^*$ that specifies the order in which the vertices of $\Gamma$ occur when traversing $Q$ from $u'$ to $v'$ (as captured by point $2$), and for each such ordering the mapping $\phi$ also determines the $Q$-distances between consecutive images of $U^*$ (as captured by point $3$, but where we only store distances of up to $\ell$ and otherwise simply mark these as \texttt{large}). This yields a unique \emph{signature} $((U^*, U_1,\dots,U_i),\lambda,\texttt{U-dist}_{\cdot,\cdot})$ for each $(\Gamma,(\phi,\psi))$. 

To determine whether $(\Gamma,(\phi,\psi))$ exists, it is sufficient to branch over all choices of signatures $((U^*, U_1,\dots,U_i),\lambda,\texttt{U-dist}_{\cdot,\cdot})$ and in each branch test for a tuple $(\Gamma,(\phi,\psi))$ which has the chosen signature. The number of signatures is upper-bounded by $p^p\cdot p^p \cdot (\ell+1)^p\in \ell^{\bigoh(\ell)}$, which also bounds the time required to enumerate these; hence, it remains to show how to determine whether there is a $(\Gamma,(\phi,\psi))$ satisfying a specified signature. Towards this, our arguments in the first paragraph directly imply the following claim:

\begin{claim*}
Let $(\Gamma,(\phi,\psi))$ satisfy the conditions of the lemma and have signature $((U^*, U_1,\dots,U_i),$ $\lambda,\texttt{U-dist}_{\cdot,\cdot})$. Then for each pair $(a,b)\in U^*$ such that $\texttt{U-dist}_{a,b}=\texttt{large}$, $\psi(ab)$ is a subpath in $Q$ whose vertices separate $\Gamma$ into a connected subgraph $\Gamma_u$ containing $u$ and $u'$, and a connected subgraph $\Gamma_v$ containing $v$ and $v'$. 
\end{claim*}

As a consequence, we may immediately discard signatures containing an edge $ab$ such that $\texttt{U-dist}_{a,b}=\texttt{large}$ and yet the edge is not an $u$-$v$ cut in $U$. As a consequence, deleting every edge $ab$ such that $\texttt{U-dist}_{a,b}=\texttt{large}$ partitions $U$ into vertex-disjoint subgraphs---each containing at least one vertex in $U^*$. Let us denote these subgraphs $U^\triangle_1,\dots, U^\triangle_j$, and note that these are uniquely defined by the signature. Moreover, since the existence of a tuple $(\Gamma,(\phi,\psi))$ satisfying a specified signature requires that $\lambda$ induces an ordering of these subgraphs, we proceed by assuming that every vertex in $U^\triangle_1\cap U^*$ occurs before every vertex in $U^\triangle_2\cap U^*$ in $\lambda$, and so forth until $U^\triangle_j\cap U^*$ (and discard our signature if this condition is not satisfied).

At this point, we are left with a signature which induces a vertex-partitioning $U^\triangle_1,\dots, U^\triangle_j$ such that a hypothetical $\Gamma$ must consist of the realizations of $(\phi,\psi)$ on each $U^\triangle_1$, \dots, $U^\triangle_j$, where each pair of consecutive realizations is separated by a path on $Q$ of length at least $\ell+1$. Note that our signature specifies the precise subgraphs $\Gamma_1,\dots,\Gamma_j$ of $G$ that form the realizations of $U^\triangle_1,\dots,U^\triangle_j$, respectively, and each such subgraph has order upper-bounded by $p+p\ell$ (it may contain only $p$ direct images of $\phi$ plus at most $p$ paths, each of length at most $\ell$ as specified by $\texttt{U-dist}$). Let $\lambda(\Gamma_1)$ be the vertex of $\Gamma_1$ which occurs on $Q$ and has minimum $Q$-distance to $u'$. We call a tuple $(\Gamma,(\phi,\psi))$ satisfying the given signature \emph{leftmost} if among all such tuples satisfying the given signature, $\Gamma$ has minimum $Q$-distance between between $u$ and $\Gamma_1$, and then among all such tuples with the same $Q$-distance between $u$ and $\Gamma_1$, $\Gamma$ has the minimum $Q$-distance between $\Gamma_1$ and $\Gamma_2$, and so forth. Clearly, if a tuple $(\Gamma,(\phi,\psi))$ exists, then so does at least one which is leftmost.

To complete the proof of the lemma, we provide a procedure to test for the existence of a leftmost tuple $(\Gamma,(\phi,\psi))$ satisfying each signature which was not discarded based on the previous two paragraphs. The procedure processes each vertex $x$ on $Q$ starting from $u$ and tests whether $G_x$ contains $\Gamma_1$ as a rooted subgraph, where $G_x$ is the subgraph of $G$ obtained after deleting the vertices of $Q$ between $u$ and $x$. Since $|V(\Gamma_1)|\leq 2\ell+2\ell^2$ and $\Gamma_1$ is planar, we can perform this test in time at most $2^{\bigoh(\ell^2)}\cdot |V(G)|$ using, e.g., the algorithm of Dorn~\cite{Dorn10}\footnote{We remark that the original version of Dorn's algorithm assumes unlabeled (i.e., non-rooted) graphs. However, it is easy to represent the roots by, e.g., subdividing each edge of $\Gamma_1$ and $G_X$ and attaching cycles of odd length to each original vertex.}. Once we find an $x$ where the test successfully finds $\Gamma_1$, we move forward along $Q$ by $\ell+1$ vertices and proceed by testing each vertex for the presence of $\Gamma_2$, and so forth until we either reach $v$ without successfully completing the test for $\Gamma_j$ (in which case we reject the current signature), or until we successfully complete the test for $\Gamma_j$ at some point before completing our processing of $v$. In the former case, we proceed to the next signature (and reject if this is the last signature to be considered); in the latter case, we output the graph $\Gamma$ consisting of the union of the individual found subgraphs $\Gamma_1,\dots,\Gamma_j$ along with the subpaths of $Q$ connecting these to $u$, $v$ and each other, along with the choice of $(\phi,\psi)$ arising from the chosen signature.

For correctness, first observe that our construction of $\Gamma_1,\dots,\Gamma_j$ from the signature guarantees that any tuple $(\Gamma,(\phi,\psi))$ we output will satisfy the conditions of the lemma. For the nontrivial direction, assume that such a tuple $(\Gamma,(\phi,\psi))$ exists in $G$. Then this tuple is associated with a unique signature $\iota=((U^*, U_1,\dots,U_i),\lambda,\texttt{U-dist}_{\cdot,\cdot})$, and there must hence exist some tuple $(\Gamma',(\phi',\psi'))$ which also satisfies the conditions of the lemma and is furthermore leftmost for $\iota$. Let $\Gamma_1,\dots,\Gamma_j$ be the subgraphs of $\Gamma'$ defined w.r.t.\ $\iota$ as above. Now each such $\Gamma_z$, $z\in [j]$, contains a vertex $x_z$ on $Q$ which is closest to $u$, and for each such vertex the subgraph isomorphism test described in the previous sentence is guaranteed to succeed. Thus, in the branch where we consider $\iota$ we are guaranteed to find some leftmost subgraph $\Gamma''$ which satisfies the conditions of the lemma, as desired.
\end{proof} \fi

We call the subgraph $\Gamma$ identified above a \emph{host} for the \emph{roadmap} $U$.

\begin{redrule}
\label{templatereduction}
Let $Q$ be a given $\ell$-clean $u'$-$v'$ path. There is a reduction rule which enumerates:
\begin{itemize}
%\item $Q$ be an $\ell$-clean $u'$-$v'$ path, 
\item the family $\mathcal{U}$ of all connected rooted graphs (with roots $u$, $v$ and \texttt{occupied}) containing at most $(2\ell^2)$-many vertices, and 
\item the set $Z$ of all vertex pairs $(u,v)$ at distance at most $\ell^2+1$ from the respective endpoints of~$Q$,
\end{itemize}
and then for each choice of $U\in \mathcal{U}$ and $(u,v)\in Z$, applies Lemma~\ref{lem:test} to determine whether there is a host of $U$ w.r.t.\ $(u,v)$ on $Q$. If there is, we \emph{mark} every vertex in $\phi(U)\cap V(Q)$ where $(\phi,\psi)$ is the computed realization.  
 
Afterwards, if there is an edge $e$ on $Q$ with two unmarked endpoints, we contract $e$.
\end{redrule}

\begin{lemma}
 
\label{lem:reductionsafe}
Reduction Rule~\ref{templatereduction} is safe. Moreover, for each $\ell$-clean path $Q$, it can be executed in time at most $2^{\bigoh(\ell^2)}\cdot |V(G)|\cdot |V(Q)|$, and is guaranteed to perform a contraction if $Q$ contains more that $32 \ell^6 \cdot 2^{14 \ell^2}+1$ 
many vertices. 
\end{lemma}
\iflong \begin{proof}
If there is a solution that does not intersect $V(Q)$, then contracting an edge in $Q$ does not change this solution. Moreover, if $\III$ does not admit a solution, then it is easy to see that the instance obtained by contracting an arbitrary edge between two free vertices does not admit a solution either. Hence, to show that the reduction rule is safe, we can assume that $\III$ admits a solution that intersects $Q$. 

Since $Q$ is $\ell$-clean, it follows from Definition~\ref{def:clean} that there exists a solution $\SSS$ and vertices $u$ and $v$ at $Q$-distance at most $\ell^2+1$ from $u'$ and $v'$, respectively, such that $\{u,v\}$ is a cut between $\{s,t\}$ and the $u$-$v$ subpath of $Q$. Moreover, by Lemma~\ref{lem:pendingvertices}, we can assume that $\SSS$ is $(Q,u,v)$-canonical. Hence, by Lemma~\ref{lem:template}, the subgraph $Q^{\SSS}_{uv}$ of $G_{\SSS}$ is a topological minor model of a graph $U\in \mathcal{U}$ such that all waiting vertices of $\SSS$ are in $\phi(U)$ and, for every $e'\in E(U)$, either $\psi(e')$ is an edge in $G$ or a subpath of $Q$. 
Furthermore, note that since $\phi(U)$ contains all the waiting vertices (and hence also all the important vertices) in $Q_{uv}^{\SSS}$, $U$ is a subgraph of a representation $H_{\SSS}$ of $\SSS$, which is a minor of $G_{\SSS}$ and the vertices $\{u,v\}$ are also a cut in $H_{\SSS}$. By Lemma~\ref{lem:equivalence}, we only need to argue that after contracting the edge $e$, we preserve at least one topological minor model of $H_{\SSS}$. 
First let us notice that $e$ has to be between $u$ and $v$ on $Q$, as for each pair $(u^*,v^*)\in Z$, the subpath of $Q$ between $u^*$ and $v^*$ is the host of the graph consisting of the edge $u^*v^*$ on $Q$ and hence all the vertices at distance at most $\ell^2+1$ from $u'$ or $v'$ are marked. 
By the choice of the edge $e$, neither of the endpoints of $e$ is in $\phi(U)$. If there is $e'\in E(U)$ such that $\psi(e')$ intersects $e$, then since $\psi(e')$ is a subpath of $Q$ between two vertices in $\phi(U)$, it follows that $e$ is an edge in the path $\psi(e')$, and contracting $e$ on $\psi(e')$ results in a path between the same two endpoints. Hence, the graph obtained by performing a contraction of edge $e$ still contains a topological minor model of $H_{\SSS}$ and a solution.

To derive the upper bound on the execution time of the reduction rule, note first that the number of pairs $(u, v)$ to be enumerated is at most $(\ell^2+1)^2$. Second, the number of rooted connected planar graphs on $q \leq 2\ell^2$ vertices is at most $4^{q} \cdot 2^{5q}=2^{7q}$. The aforementioned upper bound follows from the fact that each of the $q$ vertices is assigned one of the four possible labels, and there are at most $2^{5q}$ many connected planar graphs on $q$ vertices~\cite{BonichonGHPS06}. It follows that the number of connected rooted planar graphs on at most $2\ell^2$ vertices is at most $\sum_{q=1}^{2\ell^2}2^{7q} \le 2^{14 \ell^2+1}$.

For a fixed such labeled graph $U$ and a pair $(u, v)$, checking whether $U$ has a host takes $2^{\Oh(\ell^2)} \cdot |V(G)| \cdot |V(Q)|$ time by Lemma~\ref{lem:test}. It follows from the above that reduction rule~\ref{templatereduction} can be executed in time  $2^{\bigoh(\ell^2)}\cdot |V(G)|\cdot |V(Q)|$. 

Since for each pair $(u, v)$ the number of labeled graphs on $2\ell^2$ vertices that are enumerated is $2^{14 \ell^2+1}$, and since each such enumeration results in marking at most $2 \ell^2$ vertices of $Q$, overall, the process results in marking at most $(\ell^2+1)^2 \cdot (2 \ell^2) \cdot 2^{14 \ell^2+1} \leq 16 \ell^6 \cdot 2^{14 \ell^2}$ many vertices of $Q$. It follows that if $Q$ contains more than $32 \ell^6 \cdot 2^{14 \ell^2}+1$ vertices, then $Q$ contains two consecutive unmarked vertices, to which the reduction rule applies. 
\end{proof} \fi

\begin{theorem}
\label{thm:ellmain}
\planaroneagentproblem is fixed-parameter tractable parameterized by $\ell$.
\end{theorem}

\begin{proof}
We begin by applying Lemma~\ref{lem:freepath} to each free component in $G$ of size at least $\big(g(\ell)+1) \cdot (\ell-1)\big)+3(\ell+2)$. If this solves the instance, we are done; otherwise, it is guaranteed to either produce an equivalent instance obtained by an edge contraction (in which case we repeat), or identify an $\ell$-clean path $Q$ of length more than $g(\ell)= 32 \ell^6 \cdot 2^{14 \ell^2}+1$. In the third case, we apply Reduction Rule~\ref{templatereduction} to $Q$ to produce an equivalent instance obtained by an edge contraction in $Q$---as guaranteed by Lemma~\ref{lem:reductionsafe}---and restart.

The exhaustive application of the above procedure either correctly solves the input instance, or results in an equivalent instance such that each free component in $G$ has size at most $\big(g(\ell)+1) \cdot (\ell-1)\big)+3(\ell+2)-1$. At this point, we invoke Lemma~\ref{lem:smallfree} to solve the instance. The overall running time is upper-bounded by $2^{\bigoh(\ell^2)}\cdot |\III|^{\bigoh(1)}+\zeta(\ell)\cdot |\III|$, where the former term is the running time of the preprocessing steps and $\zeta$ is a computable function that results from the application of Lemma~\ref{lem:smallfree} (and specifically of Courcelle's Theorem within that lemma) on an instance with the aforementioned bound on the size of free components.
%free component of size at least *** in $G$ until the rule is no longer applicable to any such component. Each application reduces the number of vertices in the graph, and hence there can be at most $|V(G)|$ many applications. Afterwards, each free component in the graph has size at most ***. The result now follows from Lemma~\ref{lem:smallfree}. 
\end{proof} 

\section{Conclusion}
\label{sec:conclusion}
In this paper, we designed \FPT{} algorithms for two important variants of the Coordinated Motion Planning problem on undirected graphs, CSMP and Planar CSMP-1, in which robots move by sliding along unobstructed paths. Our results take a different perspective and contribute new insights to the design of \FPT{} algorithms for coordinated motion planning problems on graphs, developing and employing new techniques that combine topological graph theory, fixed-parameter tractability and computational logic.  
Our work opens up several interesting questions pertaining to the parameterization by the makespan $\ell$.

  \begin{enumerate}
 \item Is \cmpl\ parameterized by the makespan \W[1]-hard on general graphs?
 \item Can Theorem~\ref{thm:ellmain} be generalized to solve \cmpl\ on planar graphs?
 \item What is the parameterized complexity of CSMP and CSMP-1 on grids where each sliding-move/path is either horizontal or vertical? 
\end{enumerate}

\bibliographystyle{plain}
\bibliography{ref}
\end{document}